\documentclass{aa}

\usepackage{ulem}
\usepackage[T1]{fontenc}
\usepackage{textcomp}
\usepackage{gensymb}
\usepackage{array}
\usepackage{fixltx2e}
\usepackage{graphicx}
\usepackage[english]{babel}
\usepackage{blindtext}
\usepackage{lipsum}
\usepackage{color}
\usepackage{verbatim}
\usepackage{makecell}
\usepackage{xspace}
\usepackage{amsmath}
\usepackage{amssymb}
\usepackage{longtable}
\usepackage{multirow}
\usepackage{txfonts}
\usepackage{array}
\usepackage{bmpsize}
\usepackage{rotating}
\usepackage{ifthen}
\usepackage{fixltx2e}
\usepackage{lscape}
\usepackage{booktabs}
\usepackage{float}
\usepackage{mdframed}
\usepackage{soul}
\usepackage{lipsum}

\begin{document} 

   \title{Identifying frequency decorrelated dust residuals in B-mode maps by exploiting the spectral capability of bolometric interferometry}
   \titlerunning{Identifying frequency decorrelated dust residuals in B-mode maps}

    \author{M.~Regnier\inst{1}
        \and
        E.~Manzan\inst{2,3}
        \and
        J-Ch.~Hamilton\inst{1}
        \and
        A.~Mennella\inst{2,3}
        \and
        J.~Errard\inst{1}
        \and
        L.~Zapelli\inst{2,3}
        \and
        S.A.~Torchinsky\inst{1,4}
        \and
        S.~Paradiso\inst{5,6}
        \and
        E.~Battistelli \inst{8}
        \and
        P.~De Bernardis\inst{8}
        \and
        L.~Colombo \inst{2}
        \and
        M.~De Petris \inst{8}
        \and
        G.~D’Alessandro \inst{8}
        \and
        B.~Garcia \inst{11}
        \and
        M.~Gervasi \inst{10}
        \and
        S.~Masi \inst{8}
        \and
        L.~Mousset \inst{7}
        \and
        N.~Miron Granese \inst{13, 14, 15}
        \and
        C.~O’Sullivan \inst{9}
        \and
        M.~Piat \inst{1}
        \and
        E.~Rasztocky \inst{12}
        \and
        G.E~Romero \inst{12}
        \and
        C.G.~Scoccola \inst{13, 14}
        \and
        M.~Zannoni \inst{10}}

\institute{Université Paris Cité, CNRS, Astroparticule et Cosmologie, F-75013 Paris, France
        \and
            Università degli studi di Milano, Italy
        \and 
            INFN sezione di Milano, 20133 Milano, Italy
        \and
            Université PSL, Observatoire de Paris, AstroParticule et Cosmologie, F-75013 Paris, France
        \and
            Waterloo Centre for Astrophysics, University of Waterloo, Waterloo, ON N2L 3G1, Canada
        \and
            Department of Physics and Astronomy, University of Waterloo, Waterloo, ON N2L 3G1, Canada
        \and
            Institut de Recherche en Astrophysique et Planetologie, Toulouse (CNRS-INSU), France
        \and
            Universita di Roma - La Sapienza, Italy
        \and
            National University of Ireland, Maynooth, Ireland
        \and
            Università di Milano – Bicocca and INFN Milano-Bicocca
        \and
            ITeDA-Mza.(CNEA, CONICET, UNSAM)
        \and
            Instituto Argentino de Radioastronomía (CCT La Plata, CONICET; CICPBA; UNLP), Buenos Aires, Argentina
        \and
            Consejo Nacional de Investigaciones Científicas y Técnicas (CONICET), Godoy Cruz 2290, Ciudad de Buenos Aires C1425FQB, Argentina
        \and
            Facultad de Ciencias Astronómicas y Geofísicas, Universidad Nacional de La Plata, Paseo del Bosque, La Plata B1900FWA, Buenos Aires, Argentina
        \and 
            Universidad de Buenos Aires, Facultad de Ciencias Exactas y Naturales, Departamento de Física, Intendente Güiraldes 2160, Ciudad Universitaria, Ciudad de Buenos Aires C1428EGA, Argentina
}

   \date{Received September 15, 1896; accepted March 16, 2197}

  \abstract
    {Astrophysical polarized foregrounds represent the most critical challenge in Cosmic Microwave Background (CMB) $B$-mode experiments, requiring multi-frequency observations to constrain astrophysical foregrounds and isolate the CMB signal. However, recent observations indicate that foreground emission may be more complex than anticipated. Not properly accounting for these complexities during component separation can lead to a bias in the recovered tensor-to-scalar ratio.}  
    {In this paper we investigate how the increased spectral resolution provided by band-splitting in bolometric interferometry (BI) through a technique called \textit{spectral imaging} can help control the foreground contamination in the case of unaccounted Galactic dust frequency decorrelation along the line-of-sight (LOS).} 
    {We focus on the next-generation ground-based CMB experiment CMB-S4 and compare its anticipated sensitivity, frequency, and sky coverage with a hypothetical version of the same experiment based on bolometric interferometry (CMB-S4/BI). We perform a Monte-Carlo analysis based on parametric component separation methods (FGBuster and Commander) and compute the likelihood of the recovered tensor-to-scalar ratio, $r$.}
    {The main result is that spectral imaging allows us to detect systematic uncertainties on $r$ from frequency decorrelation when this effect is not accounted for in component separation. Conversely, an imager like CMB-S4 would detect a biased value of $r$ and would be unable to spot the presence of a systematic effect. We find a similar result in the reconstruction of the dust spectral index, where we show that with BI we can measure more precisely the dust spectral index also when frequency decorrelation is present and not accounted for in component separation.} 
    {The in-band frequency resolution provided by BI allows us to identify dust LOS frequency decorrelation residuals where an imager of similar performance would fail. This opens the prospect of exploiting this potential in the context of future CMB polarization experiments that will be challenged by complex foregrounds in their quest for $B$-modes detection.}


   \maketitle
%

\section{Introduction}
\label{sec_introduction}


This paper addresses one of the burning questions currently concerning the CMB community: Are there reliable strategies to validate or invalidate a detection of primordial \textit{B}-modes in the presence of complex, polarized Galactic foregrounds? The scope of our paper is to investigate a possible solution that exploits the spectral imaging capability of an unconventional technique for CMB polarimetry, called bolometric interferometry (BI), applied to control interstellar dust foreground emission residuals. 

Indeed, the next generation of satellites, like Litebird \citep{Hazumi_2019} and PICO \citep{hanany2019pico}, and ground-based experiments, like Simons Observatory \citep{Ade_2019} and CMB-S4 \citep{Abazajian_2022}, aim at improving the constraint on the tensor-to-scalar ratio, $r$, at the level of 0.001 and below. The accurate removal of foreground and instrumental systematic effects is already the main limiting factor.

To improve foreground removal, modern experiments are relying on multi-frequency observations and improved models of astrophysical emissions. For example, there are many PySM\footnote{\url{https://pysm3.readthedocs.io/en/latest/}} \citep{Thorne_2017} models that have been developed with the goal of simulating the effects of deviations from the single modified blackbody (MBB) emission conventionally assumed for the Galactic dust thermal emission.  The models \textbf{d5} and \textbf{d7} take into account different dust grain compositions \citep{Hensley_2017}, while the models \textbf{d4} and \textbf{d12} describe the dust emission as a sum of two or up to six single MBBs along each line-of-sight (LOS) \citep{Finkbeiner_1999, Martinez_Solaeche_2018}.

This article focuses on the \textbf{d6} model \citep{Vansyngel_2018}, which introduces LOS frequency decorrelation due to a frequency-varying polarization angle, which in turn is caused by a change both in the spectral energy distribution (SED) and in the magnetic field orientation along the LOS~\citep{tassis2015searching}.

This effect is usually quantified at the power spectra level by means of the correlation ratio, $R_{\ell}$, between two frequency maps \citep{planck_2017_decorrelation}. The most recent observational evidence regarding this effect comes from \citet{planck_2017_decorrelation, planck_2020_decorrelation, Pelgrims_2021, Ritacco_2023} and could affect polarimetric and spectral calibration in the case of wide beam instruments \citep{Masi_2021} as well as cause a bias on the tensor-to-scalar ratio \citep{McBride_2023, Hensley_2018}.

However, the \textbf{d6} model mimics the effect of a frequency-varying polarization angle, without making any physical assumptions on the misalignment of the underlying magnetic field, by randomly sampling a frequency-varying multiplication factor from a gaussian distribution, that is later applied to the single MBB emission, using the parametric expression of the correlation ratio derived in \citet{Vansyngel_2018}.



If dust does not behave as a simple MBB, as is usually assumed, but exhibits more complex spectral features, like frequency decorrelation, we need a method to detect the presence of foreground residuals in our results.
This could be achieved by comparing results from different sky patches, as proposed by \citet{aurlien2022foreground}, or by cross-checking with different component separation methods, such as parametric codes \citep{eriksen_2006, Stompor}, blind algorithms \citep{Aumont_2007} or codes based on moment expansion \citep{Chluba_2017, Vacher_2022}, some of which might be less sensitive to foreground incorrect modeling.

Another possibility, which we illustrate in this paper, is to use BI and its ability to discriminate frequencies in-band during data analysis. This allows us to achieve a spectral resolution of a few~GHz\footnote{Some level of in-band frequency sensitivity is actually achievable to traditional imagers by using the small variations in the spectral properties of different detectors. This was successfully applied to map the CO emission line~\citep{Planck_CO}.} and reanalyze the same data with different spectral configurations.  A variation in the constraint on $r$ between configurations suggests contamination in the tensor-to-scalar ratio due to component separation residuals.


In this paper, we investigate the advantage of BI for foreground removal and characterization by comparing the performance in detecting dust frequency decorrelation of one of the most advanced experiments to come, CMB-S4, with a similar, hypothetical experiment based on bolometric interferometry, that we name CMB-S4/BI. We perform a Monte-Carlo simulation starting from frequency maps, with or without band-splitting, and then we apply parametric component separation using two different component separation codes: FGBuster \citep{Stompor} and Commander \citep{eriksen_2006,eriksen_2008}. In the main body of this paper, we focus on FGBuster simulations, while we discuss the results obtained with Commander in Appendix~\ref{app:commander}.

Because the aim of this article is to propose a new methodology, we did not perform an actual map-making process from the Time-Ordered-Data, but we simulated the noise properties directly onto the reconstructed frequency maps, and we neglected the impact of instrumental systematic effects, such as: an imperfect knowledge of the spectral response of the instrument, an uncertainty about the Half-Wave Plate angle, or the feed-horn positions. 
Such effects could reduce the ability to perform band-splitting during data analysis. However, BI offers a specific approach to control instrumental systematic effects, the \textit{self-calibration} technique, that is inherited from radio-interferometers ~\citep{Bigot-Sazy2013}.

The paper is organized as follows. In Sect.~\ref{sec_bolometric_interferometry_and_qubic} we provide a brief introduction to BI \citep[and references therein]{2020.QUBIC.PAPER1}. 
Sect.~\ref{sec_methods} is dedicated to the description of the simulated sky models, instrumental configurations, and the Monte-Carlo pipeline based on the FGBuster \citep{Stompor} component separation code. In Sect.~\ref{sec_results} we compare the results in terms of tensor-to-scalar ratio reconstruction from simulations with conventional foreground models and with unaccounted Galactic dust LOS frequency decorrelation. Here we also describe a machine learning classification used to assess the ability to detect residuals from foreground emissions in a single realization. In Appendix~\ref{app:reconstruction_foregrouds} we present the results obtained with FGBuster regarding the estimation of foreground parameters, while in Appendix~\ref{app:commander} we discuss all the results obtained with Commander.  


\section{Bolometric interferometry in a nutshell}
\label{sec_bolometric_interferometry_and_qubic}

    In this section we briefly describe the principles of BI, focusing on a specific feature of this technique, called \textit{spectral imaging}, which is at the heart of our study. 
    The interested reader can find more details on BI and \textit{spectral imaging} in \cite{2020.QUBIC.PAPER1, 2020.QUBIC.PAPER2}, whereas more information about the QUBIC experiment, currently the only one based on BI, and on its laboratory characterization can be found in \cite{2020.QUBIC.PAPER3, 2020.QUBIC.PAPER4, 2020.QUBIC.PAPER5, 2020.QUBIC.PAPER6, 2020.QUBIC.PAPER7, 2020.QUBIC.PAPER8}.

    \subsection{Principles of bolometric interferometry}
    \label{sec_bolometric_interferometry}
    
        Bolometric interferometry is a technique that combines the use of bolometers, which are state-of-the-art wide-band cryogenic detectors providing high sensitivity, with the advantage of precision control of systematic effects provided by the \textit{self-calibration} technique, commonly used in radio-interferometry \citep{1981MNRAS.196.1067C}. The application to BI is detailed in \citet{Bigot-Sazy2013}.

\begin{figure}[ht]
	\includegraphics[width=9cm]{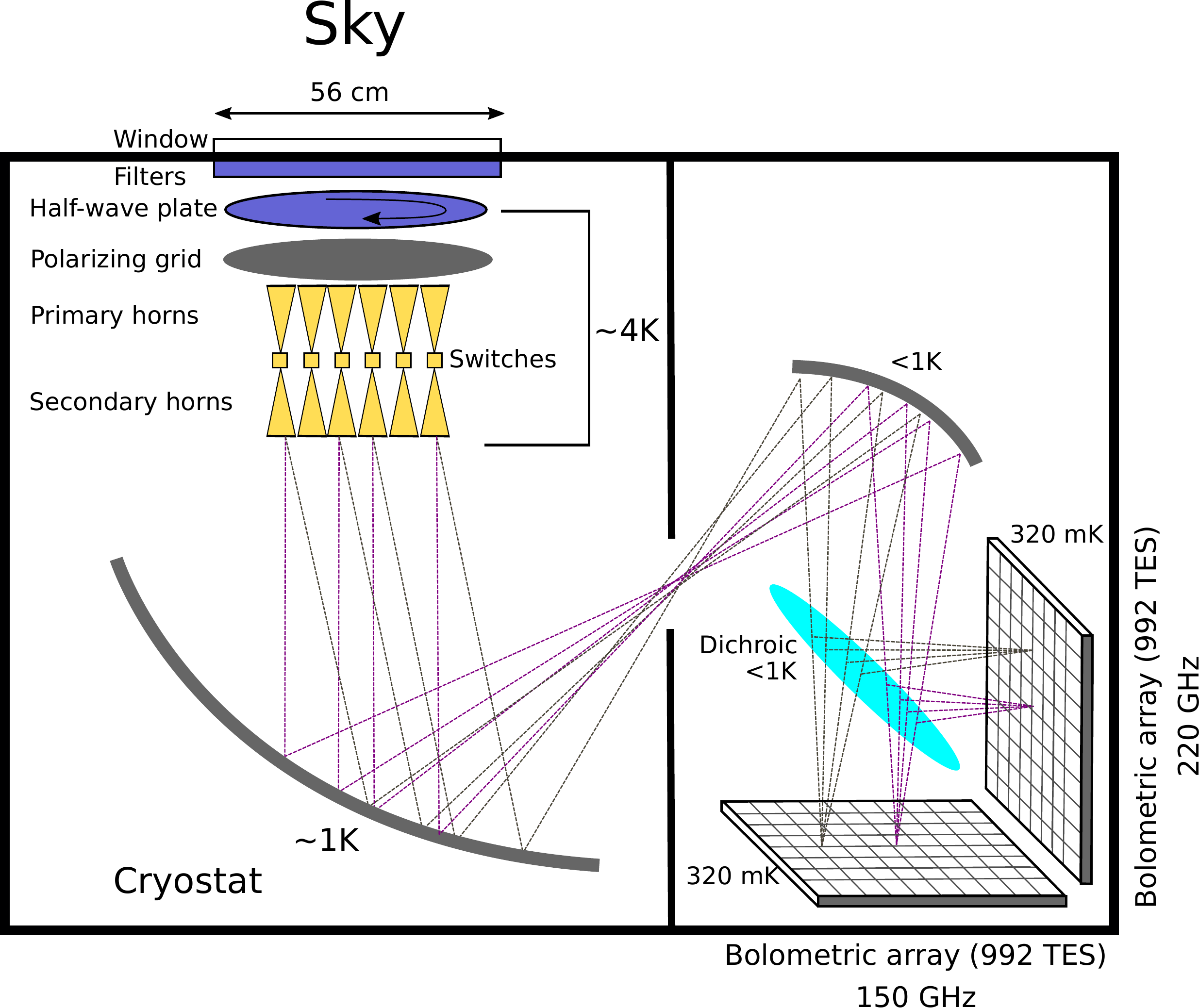}
        \caption{\label{fig:BI_instrumental_concept}Schematic of the QUBIC instrument showing the principle of bolometric interferometry. The sky signal is received by an array of back-to-back horns and re-imaged onto the bolometric focal planes where the field interferes additively. A polarizer and a rotating half-wave plate make the instrument sensitive to linear polarization.}
\end{figure}

Figure~\ref{fig:BI_instrumental_concept} shows a schematic of the QUBIC instrument, highlighting the fundamentals of BI. The sky signal enters the cryostat through an aperture window and propagates through a series of filters, a step-rotating half-wave plate, a polarizing grid, and an array of paired back-to-back feed-horn antennas. The back horns directly illuminate an optical combiner, which focuses the radiation onto two focal planes through a dichroic plate. 

\begin{figure}[ht]
    \centering
    \includegraphics[width=7cm]{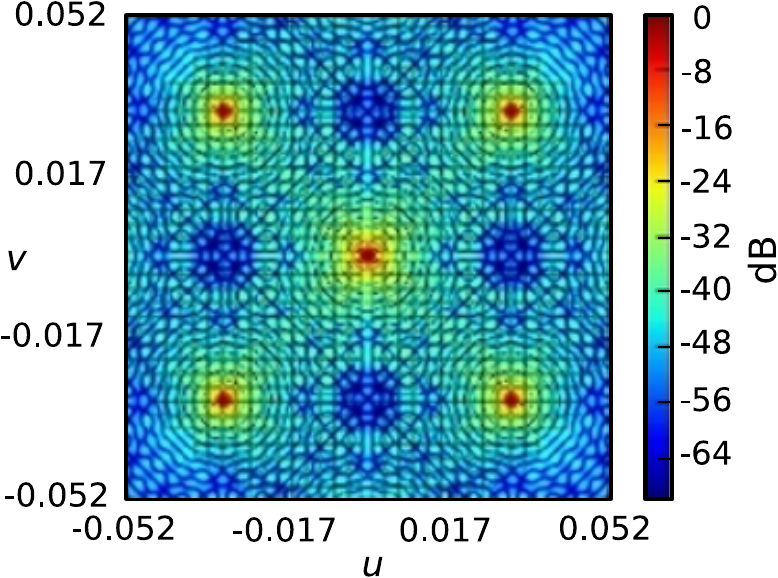} \\
    \includegraphics[width=9cm]{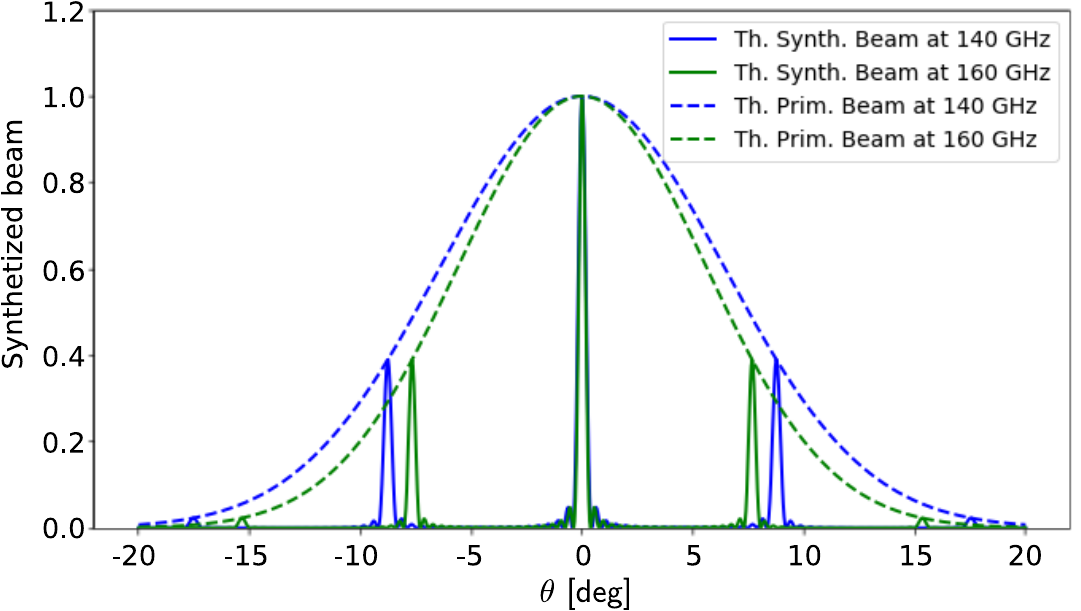}
        \caption{\label{fig:Theo_SB} \textit{\textbf{Top} panel}: simulation of the interference pattern on the focal plane generated by a monochromatic point source. \textit{\textbf{Bottom} panel}: azimuth cut of the theoretical synthesized beam (solid lines) at 140 GHz (blue line) and at 160 GHz (green line) for a detector at the center of the focal plane. Dashed lines represent the beam pattern of a single feed horn. The frequency-dependent position of the secondary peaks is clearly visible.}
\end{figure}

When the instrument observes a distant point source along the optical axis an interference pattern forms on the two focal planes (see the top panel of Fig.~\ref{fig:Theo_SB}). As a result, each focal plane element measures the sky signal convolved by a specific beam pattern, called the \textit{synthesized beam}, shown in the bottom panel of Fig.~\ref{fig:Theo_SB}. The constructive or destructive interference of the incoming signal defines a series of peaks and nulls, with properties that depend on the signal wavelength, $\lambda$, on the number of horns along the maximum axis of the antennas array, $P$, and on the separation between two consecutive horns, $\Delta h$, as follows \citep{2020.QUBIC.PAPER2}:

\begin{equation}
    \label{eq_synth_beam_properties_dependece}
    \theta_\mathrm{FWHM}  = \frac{\lambda}{(P-1)\Delta h},\,\,\,\,\,\Theta = \frac{\lambda}{\Delta h},
\end{equation}
where $\theta_\mathrm{FWHM}$ is the half power width of the peaks and $\Theta$ is the angular distance between the main peak and the first secondary peak.

Equation~\ref{eq_synth_beam_properties_dependece} demonstrates that the positions of the secondary peaks depend on $\lambda$. As an example, in the bottom panel of Fig.~\ref{fig:Theo_SB} we show a cut of the synthesized beam at a fixed azimuthal angle for two frequencies: 140\,GHz and 160\,GHz. Knowing how the multiple-peaked shape of the synthesized beam evolves with frequency allows us to recover the sky signal during data analysis at various frequencies within the physical band. This is possible as long as the two frequencies, $\nu_1$ and $\nu_2$, are far enough apart that the secondary peaks are well-resolved.  That is, we require $\Theta(\nu_2) -  \Theta(\nu_1) > \theta_\mathrm{FWHM} (\sqrt{\nu_1\nu_2})$, which occurs for $\frac{\Delta \nu}{\nu} \geq \frac{1}{P-1}$. We call this technique \textit{spectral imaging}.

Our goal is to reconstruct maps of an extended source in polarization thereby computing the three Stokes parameters I, Q, and U at the same time. Because an extended source is a linear combination of point-sources, this reconstruction is possible but requires deconvolving from the multiple peaks of the synthesized beam, as well as relying on a half-wave plate modulation for polarization reconstruction.
This problem can be solved thanks to a scanning strategy which allows information to be captured several times with various geometrical configurations\footnote{Similar to grism spectroscopy that benefits from different orientations of the field of view.}, and through an inverse problem approach that reconstructs unbiased maps of the three Stokes parameters in sub-bands within the physical band of the instrument ~\citep{2020.QUBIC.PAPER2}.





Consequently, the frequency dependence of the secondary peaks enables us to achieve a spectral resolution of a few GHz within the physical band. Furthermore, since spectral imaging occurs at the data analysis level, it allows us to re-analyze the same data with different spectral configurations, which can help us detect biases in the obtained results. This is a unique asset compared to traditional imagers, which would need several focal planes coupled to multichroic filters to achieve the same spectral performance, or to Fourier-transform spectrometers, which would suffer from a noise penalty related to not observing all frequencies simultaneously.


In this context, our aim is to investigate how the increased spectral resolution provided by BI helps in controlling the contamination from Galactic foregrounds in the quest for primordial $B$-modes detection, with a special focus on the Galactic dust emission.

\section{Dust decorrelation with bolometric interferometry and direct imaging}

\label{sec_detecting_dust_decorrelation}

    This paper aims to quantify the effect of various dust models with increasing complexity on the component separation results and demonstrate the benefits of spectral imaging in this regard. We focus, in particular, on the LOS frequency decorrelation of thermal dust, a phenomenon already observed in Planck data \citep{Pelgrims_2021}. 
    
    To quantify dust decorrelation, we follow \citet{planck_2020_decorrelation}, and use the quantity $\mathcal{R}_\ell$, defined in Eq.~(\ref{Rl}): 
    
    \begin{equation}
        \mathcal{R}_{\ell}^{\nu_1 \times \nu_2} = \frac{\mathcal{C}_{\ell}^{\nu_1 \times \nu_2}}{\sqrt{\mathcal{C}_{\ell}^{\nu_1 \times \nu_1} \times \mathcal{C}_{\ell}^{\nu_2 \times \nu_2}}}.
        \label{Rl}
    \end{equation}
    
    $\mathcal{R}_\ell$ is the ratio between the crossed spectrum between two frequencies, $\nu_1$ and $\nu_2$, and the square root of the product of the auto-spectra at these same frequencies. This ratio is close to 1 for completely correlated thermal dust. In our sky simulations, we can increase or decrease the level of complexity in the thermal dust spectral energy density (SED) by tuning $\mathcal{R}_{\ell}$ to a value farther or closer to one, thanks to the parametric expression of $\mathcal{R}_{\ell}$ derived in Eq.~(14) of \citet{Vansyngel_2018}.
    
    To assess the potential of BI, we compare the component separation performance of CMB-S4 to a BI version of the same experiment having the same sensitivity per unit bandwidth, but allowing for a higher spectral resolution through band-splitting using spectral imaging. In the following subsections, we present the methods used for this comparison.

    \subsection{Methods}
    \label{sec_methods}

        \subsubsection{Simulated sky}
        \label{sec_simulated_sky_model}
            Our sky model contains the CMB plus synchrotron and dust emission foregrounds. 

We simulated the CMB using angular power spectra provided by the \texttt{fgbuster} package that are based on the latest Planck 2018 results\footnote{Spectra can be accessed at \protect\url{https://github.com/fgbuster/fgbuster/tree/master/fgbuster/templates}}. We used the following two FITS files: 
\begin{enumerate}[(i)]
\item \texttt{\small{Cls\_Planck2018\_lensed\_scalar.fits}} in which $B$-modes are considered with $r=0$ and lensing, 
\item \texttt{\small{Cls\_Planck2018\_unlensed\_scalar\_and\_tensor\_r1.fits}} in which $B$-modes are considered with $r=1$ and no lensing. 
\end{enumerate}

In our simulations, we used $TT$, $EE$, and $TE$ spectra taken directly from the file (i), while the $BB$ spectrum was obtained by summing the $BB$ spectrum from the file (i) multiplied by a lensing residual of 0.1 with the $BB$ spectrum from the file (ii) multiplied by the value of $r$ (either 0 or 0.006). Note that such a simplified approach neglects the additional tensor contribution to the $TT$, $TE$, and $EE$ spectra, but is sufficient in our case, as we only perform the likelihood analysis on the $BB$ spectrum.


For the foregrounds we considered the following models\footnote{See \protect\url{https://pysm3.readthedocs.io/en/latest/#models} for more details about the models.}:

%
%
%

\begin{enumerate}
    \item model \textbf{d0s0}, which assumes a single modified black-body (MBB) emission for the thermal dust and a power-law emission for the synchrotron with no curvature, with constant dust spectral index across the sky, $\beta_\mathrm{d} = 1.54$, dust temperature, $T_\mathrm{d} = 20$\,K, and synchrotron spectral index, $\beta_\mathrm{s} = -3$;

    \item model \textbf{d1s1}, derived from the Planck data post-processed with the Commander code \citep{Planck_15} for the dust emission, while the synchrotron emission is taken from the Haslam data at 408 MHz in \cite{remazeilles2015improved}, \cite{Haslam82}. The thermal dust emission is modeled as a modified black body with spatially varying temperature and spectral index projected on the sky, while the synchrotron emission is modeled as a power-law with spatially varying spectral index with no curvature;
    \item model \textbf{d6s1}. This model is derived from \textbf{d1s1} with the introduction of LOS frequency decorrelation in the dust emission following the statistical approach described in Eq.~(14) of \citet{Vansyngel_2018}. 
\end{enumerate}

\begin{figure}
    \centering
    
    \includegraphics[width=9cm]{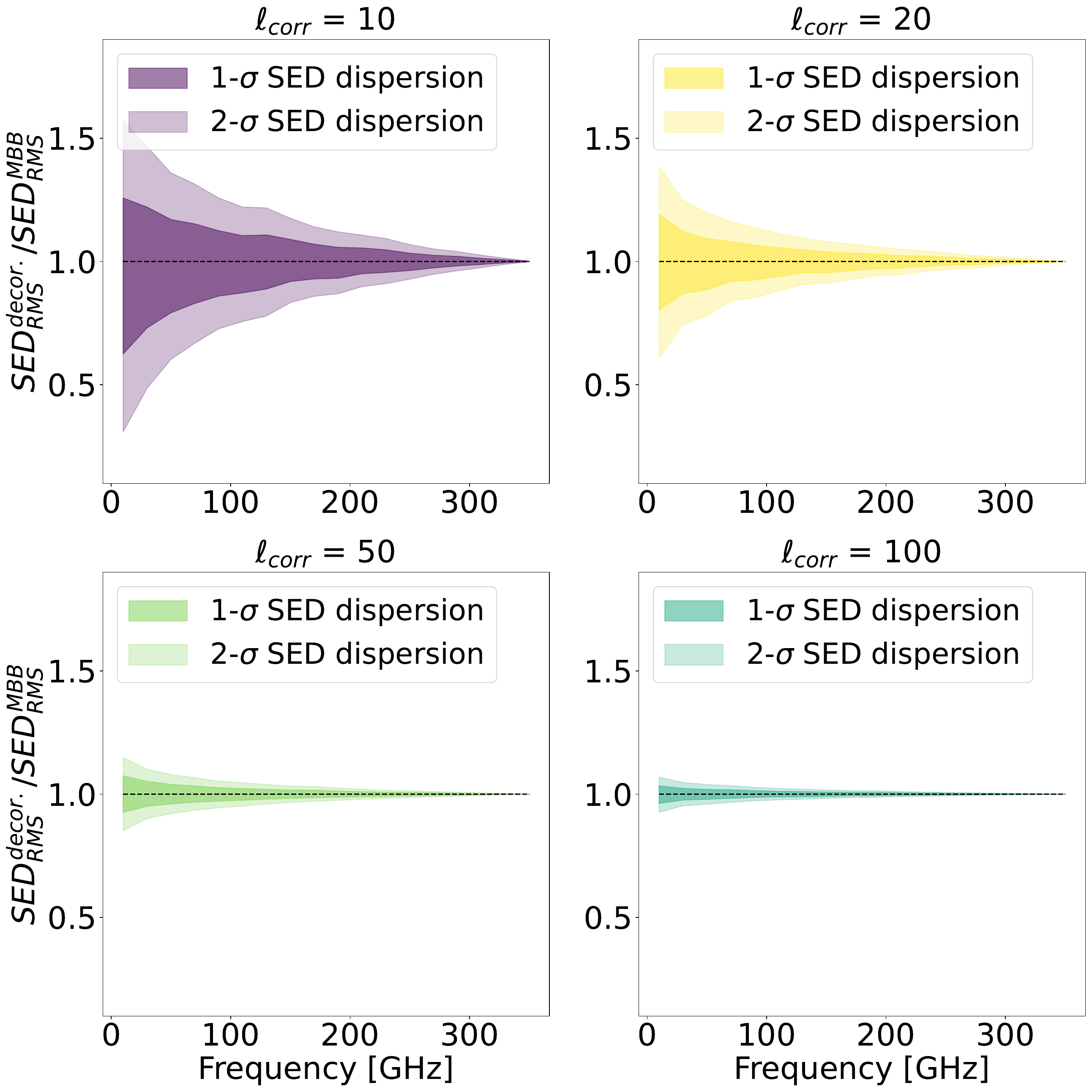}
    \caption{Dispersion of the dust SED for different correlation lengths of the PySM \textbf{d6} model normalized by the single MBB emission (\textbf{d1} model).
    The colored areas represent the statistical deviation from an MBB for a given correlation length, evaluated over 500 realizations.
    }
    \label{fig:Dust_d6_SED_dispersion}
\end{figure}

Whereas models \textbf{d0s0} and \textbf{d1s1} are fixed realizations, the model \textbf{d6s1} results in a random realization of the SED. For each simulated frequency, the MBB emission is multiplied by a randomly sampled decorrelation factor that mimics the effect of a frequency-varying polarization angle without making any physical assumptions on the underlying Galactic magnetic field. The magnitude of the decorrelation factor is governed by the correlation length, $\ell_\mathrm{corr}$, a parameter that can be set in PySM. Figure~\ref{fig:Dust_d6_SED_dispersion} displays the dispersion of various SED realizations as a function of $\ell_\mathrm{corr}$, showing that the dispersion increases with a shorter correlation length.



\begin{figure}
    \centering
    \includegraphics[width=0.5\textwidth]{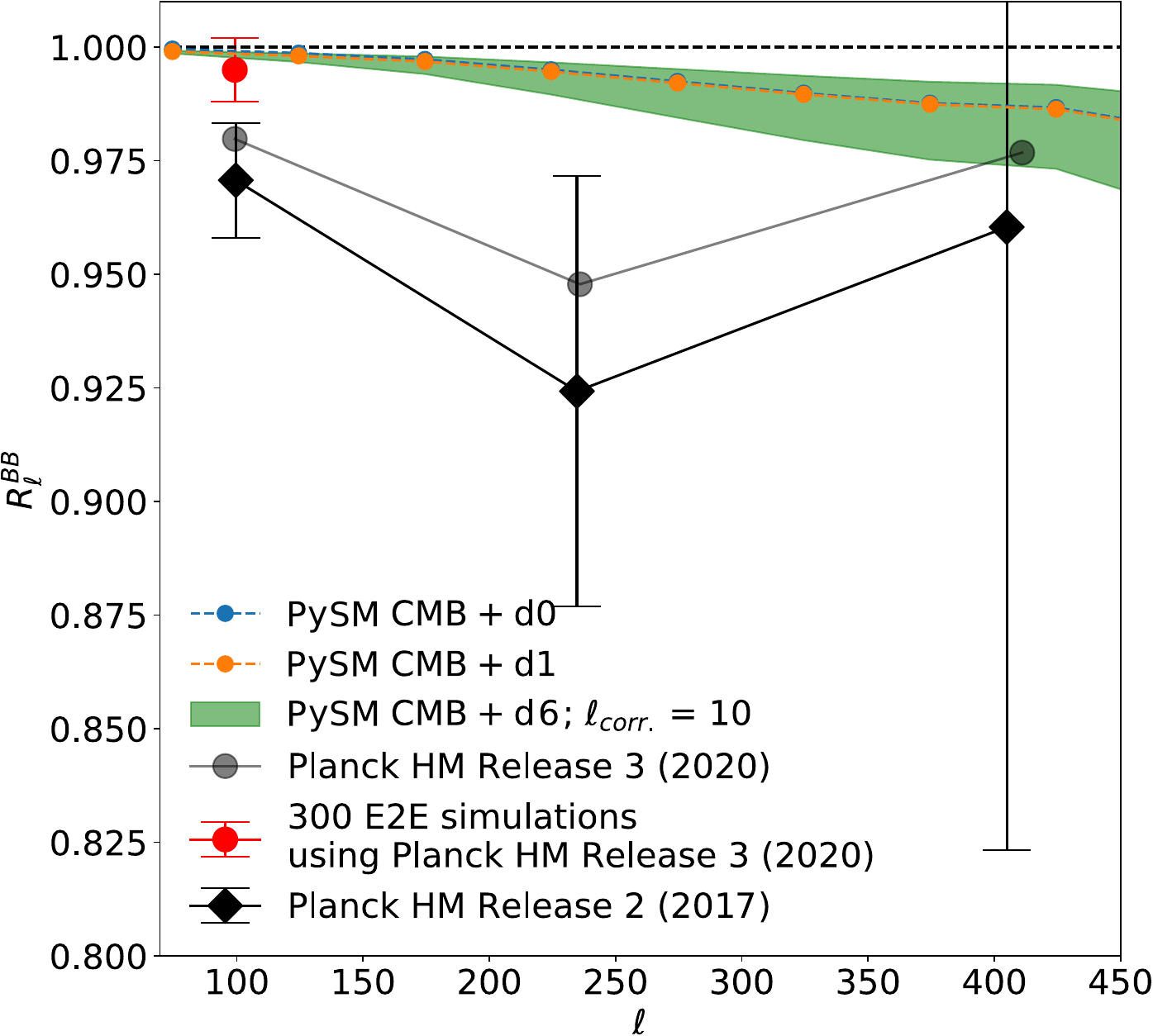}
    \caption{Correlation ratio measured by Planck from the Half Mission (HM) maps at 217 GHz and 353 GHz, compared to the simulated ratio using PySM dust and CMB templates at the same frequencies. The dots in blue and orange represent the expected $R_{\ell}$ for the CMB and a single MBB dust emission, with constant (\textbf{d0}) or varying (\textbf{d1}) spectral indices pixel-by-pixel. Note that the dots are so close that they overlap in the figure. The green envelope shows the range of $R_{\ell}$ obtained from 500 realizations of dust LOS frequency decorrelation with $\ell_\mathrm{corr} = 10$. The black dots are from Fig.~2 of \citet{planck_2017_decorrelation}, the gray dots are from Fig.~B.2 of \citet{planck_2020_decorrelation}, and the red point has been obtained from the values in the middle plot of the second row in Fig.~18 of \citet{planck_2020_decorrelation}.}
    \label{fig:Correlation_ratio_Planck_vs_pysm_models}
\end{figure}

In our simulations, we explore the effect of dust LOS frequency decorrelation with a level of decorrelation consistent with current observations. Specifically, the range of correlation lengths used in our study is $\ell_\mathrm{corr}\geq 10$, which corresponds to a decorrelation level below $5\%$ for all the simulated frequencies. This configuration represents a conservative scenario with respect to the decorrelation level measured by Planck \citep{planck_2017_decorrelation,planck_2020_decorrelation} in the same multipole range considered in our work ($\ell \leq 300$ $-$ see Fig.~\ref{fig:Correlation_ratio_Planck_vs_pysm_models} for a comparison with Planck estimates).


        \subsubsection{Instrument models}
        \label{sec_instrument_models}
            The first instrument considered in our analysis is CMB-S4 \citep{Abazajian_2022}, which will observe at 9 different frequencies in the 20--280\,GHz range to constrain both synchrotron and thermal dust emissions. The goal of CMB-S4 will be the detection of $r$ at the level $r > 0.003$ with more than 5$\sigma$. 

The second instrument is a version of CMB-S4 based on bolometric interferometry (CMB-S4/BI), where each of the bolometer-based frequency bands, $\Delta\nu_i$ (i.e. above 85 GHz), can be subdivided into $n_\mathrm{sub}$ sub-bands of width: 


\begin{equation}
    \Delta \nu^\mathrm{BI}_i = \frac{\Delta \nu_i}{n_\mathrm{sub}}.
    \label{bandwidth}
\end{equation}

If we now consider $m$ frequency bands of CMB-S4, each one subdivided in $n_\mathrm{sub}$ sub-bands in CMB-S4/BI we can calculate the sensitivity in each sub-band as:

\begin{equation}
    \sigma^\mathrm{BI}_{j,i} = \sigma_j \times \sqrt{n_\mathrm{sub}} \times \varepsilon,
    \label{sigma}
\end{equation}
where $\sigma_j$ is the CMB-S4 sensitivity in the $j$-th sub-band within $i$-th physical band, $n_\mathrm{sub}$ is the number of sub-bands and $\varepsilon$ is a parameter introduced to account for the sub-optimality of bolometric interferometry \cite[for further details about BI sub-optimality see][]{2020.QUBIC.PAPER2}.

Two approximations have been done regarding the instrument models:
\begin{enumerate}
 \item The noise is always assumed to be white, although, in \mbox{CMB-S4/BI}, we have added the multiplicative term $\varepsilon$ to account for the sub-optimality of bolometric interferometry. We know that the noise of a bolometric interferometer is not entirely white, and this calls for specific component separation techniques able to deal with correlated noise. These techniques are currently under development within the QUBIC collaboration;
 \item We have neglected the angular resolution of the optical beam to be consistent with the CMB-S4 reference paper. The angular resolution of a traditional imager, such as CMB-S4, is set by the aperture of the telescope, whereas in the BI case this is set by the largest distance between 
 horns. Although the contribution of the physical beam affects the final sensitivity of both instruments, it should not impact the generality of our results.
\end{enumerate}

\begin{figure}
    \centering
    \includegraphics[width=9cm]{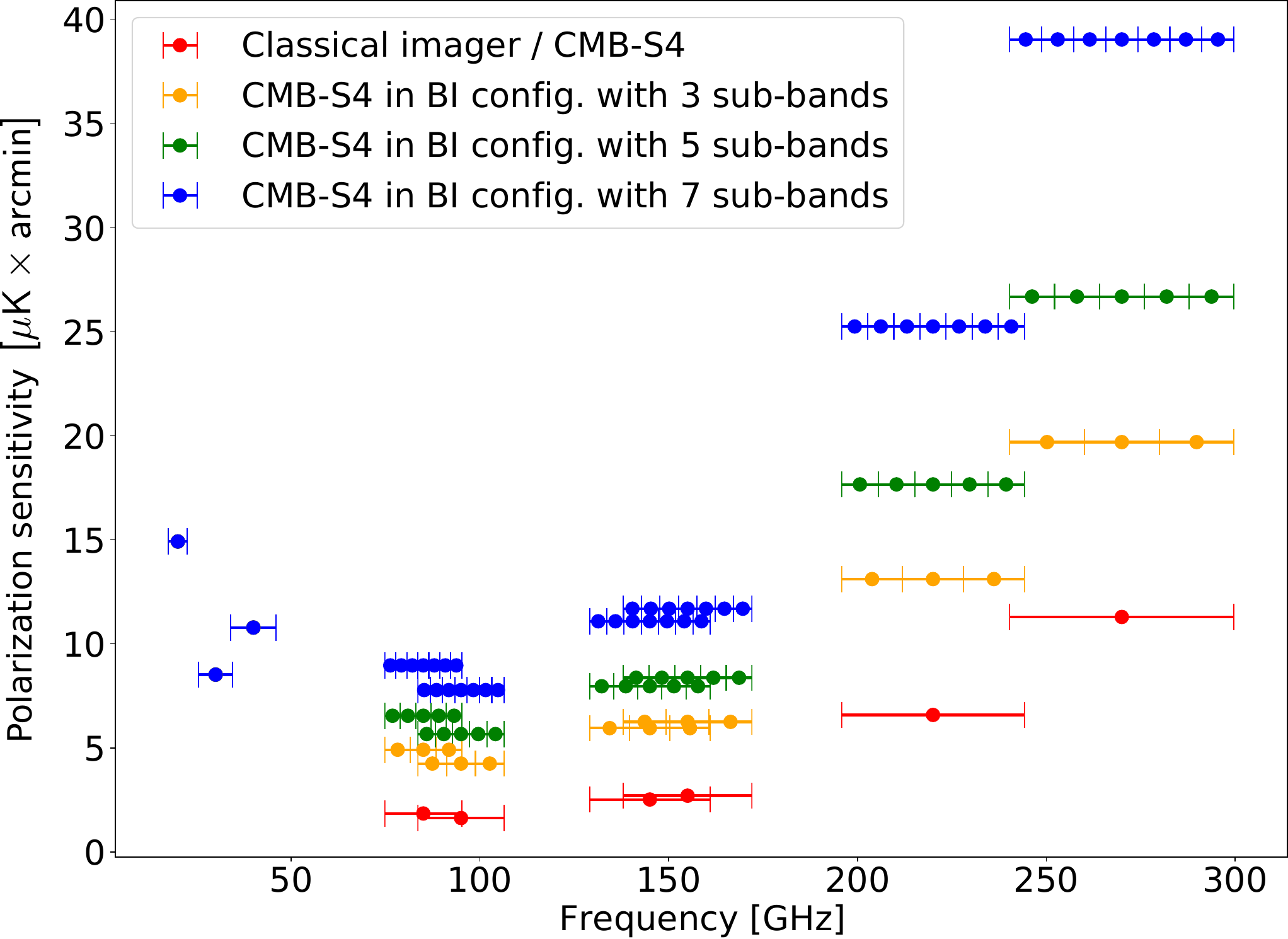}
    \caption{Polarization sensitivity of CMB-S4 and three examples of \mbox{CMB-S4/BI}, with $n_\mathrm{sub}=3,5,7$ respectively. Note that the bands of the three lowest frequency channels are identical for all the instruments. Because our study focuses on dust decorrelation we have chosen not to split the bandwidths of the synchrotron channels.}
    \label{fig:Polarization_depth}
\end{figure}

Figure~\ref{fig:Polarization_depth} shows the bandwidths and sensitivities of some of the tested experimental configurations. For each CMB-S4 frequency interval above 85\,GHz, we have studied seven configurations of CMB-S4/BI, with $n_\mathrm{sub}$ ranging from 2 to 8. Increasing the number of sub-bands results in a sensitivity degradation, as indicated in Eq.~(\ref{sigma}), with $\varepsilon$ ranging between 20\% and 60\%, according to \cite{2020.QUBIC.PAPER2}. Since we focus on dust decorrelation, we have not subdivided the synchrotron frequency bands, so that the first three intervals of the various configurations overlap. Note that because the simulated CMB-S4 sky patch is centered far away from the Galactic plane, we expect the correlations between dust and synchrotron to be negligible for the scope of our study by following \cite{Krachmalnicoff_2018} and~\cite{planck2017}.


We emphasize that because this band-splitting is performed at the data analysis level, one can explore various values of the number of sub-bands $n_\mathrm{sub}$ with the same dataset. Studying the evolution of the resulting constraints as a function of $n_\mathrm{sub}$ is the core of this study.


        \subsubsection{Simulation pipeline}
        \label{sec_simulation_pipeline}

        We describe here the simulation pipeline for the Monte-Carlo analysis that we performed using the FGBuster parametric component separation code. In Appendix~\ref{app:commander_pipeline} we report the same information regarding the simulations performed with Commander.

In the FGBuster analysis, we simulated the anticipated CMB-S4 patch, which is a 3$\%$, circular sky patch centered in RA, DEC = $(0^{\circ},-45^{\circ})$.




\begin{table}[h!]
    \caption{\label{tab:fgbuster_vs_commander} Parameters used for analyzing simulations with FGBuster for all dust models}
    \renewcommand{\arraystretch}{2}
    \begin{center}
        \begin{tabular}{p{5cm} m{3cm}}
            \rule{0.47\textwidth}{0.02cm}\\
            Map $N_\mathrm{side}$\dotfill	& \makecell[c]{256} \\
            Multipole range\dotfill	&\makecell[c]{21--335} \\
            $\Delta\ell$\dotfill &\makecell[c]{\phantom{0}{35}} \\
            Input $r$\dotfill & \makecell[c]{0, 0.006} \\
            Residual lensing fraction\tablefootmark{a}\dotfill & \makecell[c]{\phantom{000}10\%} \\
            Sky fraction [\%]\dotfill & \makecell[c]{\phantom{0000}3\%} \\
            \makecell[l]{Sky patch center\tablefootmark{b}\\ \mbox{}[Equatorial coord.]\dotfill}\dotfill & 
                \makecell[c]{$\text{RA}=0^\circ$\\\phantom{1}$\text{DEC}=-45^\circ$\rule[-2.2ex]{0pt}{0pt}}\\
            \rule{0.47\textwidth}{0.01cm}\\
            \multicolumn{2}{l}{\makecell[l]}{    
              
              
              \tablefoot{\\
                \tablefoottext{a}{This is the fraction of the lensing signal left in the CMB map.}\\
                \tablefoottext{b}{Center of the CMB-S4 sky patch.} }
            }
        \end{tabular}
    \end{center}
\end{table}




We considered eight instrument configurations (see also Fig.~\ref{fig:Polarization_depth}): The CMB-S4 configuration (parametrized following~\citet{Abazajian_2022}) and seven versions of CMB-S4/BI, obtained by subdividing each frequency band. We then applied component separation and analyzed the cross-spectra of the resulting maps using a uniform binning (see Table~\ref{tab:fgbuster_vs_commander} for a summary of the simulation set-up).

For each instrument configuration, the overall analysis chain followed these steps:
\begin{enumerate}
 \item Generate a CMB realization as described at the beginning of Sect.~\ref{sec_simulated_sky_model};
 \item Generate a noise realization for each frequency channel in the considered instrument configuration; 
 
 \item Add the CMB and the noise to the foreground maps generated as described in Sect.~\ref{sec_simulated_sky_model};
\item Apply component separation to the input maps. In some cases we assumed the same model used to generate the input case.  In others, we assumed a different model in order to mimic a realistic situation in which the actual foregrounds are not 100\% known and one might assume a model that does not completely describe reality;
 \item Perform a cross-spectra analysis between two noise realizations (each with half the exposure time) to recover the tensor-to-scalar ratio, $r$. We calculated angular power spectra using the NaMaster\footnote{\protect\url{https://namaster.readthedocs.io/en/latest/}} code \citep{Alonso_2019} with an apodization radius of 4$^\circ$.
\end{enumerate}

In Table~\ref{tab:performed_simulations} we list the various cases studied in this work. Each case was simulated with all the instrument configurations described in Sect.~\ref{sec_instrument_models} and Fig.~\ref{fig:Polarization_depth}. 

\begin{table}[h!]
    \begin{center}
        \renewcommand{\arraystretch}{1.3}
        \caption{\label{tab:performed_simulations}Cases analyzed in this work.}
        \begin{tabular}{c c}
            \hline
            {Input foreground model} & \makecell[c]{{Model assumed in} \\ {component separation}}\\
            \hline
            \hline
            \textbf{d0s0}\phantom{ ($\ell_\mathrm{corr} = \phantom{1}10$)} &\textbf{d0s0}\\
            \hline
            \textbf{d1s1}\phantom{ ($\ell_\mathrm{corr} = \phantom{1}10$)} &\textbf{d1s1}\\
            \hline
              \begin{tabular}{c r}
                 \multirow{5}{*}{\textbf{d6s1}} &\hfill $\ell_\mathrm{corr} = \phantom{1}10$ \\
                                                &\hfill $\ell_\mathrm{corr} = \phantom{1}13$ \\
                                                &\hfill $\ell_\mathrm{corr} = \phantom{1}16$ \\
                                                &\hfill $\ell_\mathrm{corr} = \phantom{1}19$ \\
                                                &\hfill $\ell_\mathrm{corr} = 100$ \\
             \end{tabular} & \textbf{d1s1}\\
            
            \hline
        \end{tabular}

    \end{center}    
\end{table}

\paragraph{Component Separation.} We performed parametric component separation modelling on our data as follows:

\begin{equation}
    \vec{d}_p = \textbf{A}\cdot \vec{s}_p + \vec{n}_p,
    \label{eq:data_model}
\end{equation}

where $p$ is the pixel index, $\vec d_p$ and $\vec{n}_p$ are vectors representing the data and noise measured by the instrument frequency channels, $\vec{s}_p$ is a vector containing the ``true'' sky values at the same frequencies, and $\textbf{A}$ is a mixing matrix that contains information about the sky components (CMB, synchrotron and interstellar dust). In our simulations, we considered the dust temperature as a known parameter, $T_\mathrm{d}=20$\,K. Thus, the only unknown parameters for synchrotron and dust emissions were their spectral indices, $\beta_\mathrm{s}$ and $\beta_\mathrm{d}$.


FGBuster solves for the best spectral indices, $\beta_\mathrm{s}$ and $\beta_\mathrm{d}$, given the data, $\vec d_p$, and the noise covariance matrix, $\textbf{N}$, following the spectral likelihood approach of \citet{Stompor}. In order to cope with computational constraints (processing time and computer memory) and keep the same parameters as in~\citet{Abazajian_2022}, we used a double pixelization scheme in our component separation: A fine resolution of $N_\mathrm{side}=256$ for the pixels of the reconstructed maps, and a coarse resolution of $N_\mathrm{side}=8$ (corresponding to a super-pixel resolution of about 7$^\circ$) for the spectral indices. In other words, the spectral indices are kept constant on larger pixels compared to those of the reconstructed maps. This approach introduces a slight bias on $r$, as demonstrated and addressed in Sect.~\ref{sec:reconstruction_r_fgbuster}. However, this bias does not alter the general validity of our results.

\paragraph{Tensor-to-scalar ratio estimation.} The main goal of our study is to assess how residuals caused by biased estimates of foreground parameters impact the reconstruction of the tensor-to-scalar ratio, $r$, which is the main parameter characterizing the primordial CMB $B$-modes. 

We write the likelihood on $r$ using a Gaussian approximation~\citep{Hamimeche_2008}:





\begin{equation}
    -2 \ln \mathcal{L}(r) = \left( \textbf{D}^{BB}_{\ell, \text{exp}} - \textbf{D}^{BB}_{\ell, \text{model}} \right)^T \textbf{N}_{\ell, \ell}^{-1} \left( \textbf{D}^{BB}_{\ell, \text{exp}} - \textbf{D}^{BB}_{\ell, \text{model}} \right),
    \label{chi2}
\end{equation}

where $\textbf{D}^{BB}_{\ell, \text{exp}}$ and $\textbf{D}^{BB}_{\ell, \text{model}}$ are the measured and theoretical angular power spectra, $\textbf{N}_{\ell, \ell}^{-1}$ is the inverse of the sum of the noise and sample variance-covariance matrices, and $\mathcal{L}(r)$ is the likelihood on $r$. The theoretical angular power spectrum, $\textbf{D}^{BB}_{\ell, \text{model}}$, includes the contribution of the $10\%$ lensing residual that we assumed throughout the study. Therefore, the only free parameter is the tensor-to-scalar ratio, $r$, which we vary with a flat prior in the range [-1, 1]. Although allowing for negative values of $r$ is unphysical, we opt for this more general approach because it has the benefit of highlighting potential biases due only to differing observational methodologies.

This work explores what happens when dust is more complex than anticipated. In order to do so, we perform component separation assuming a simple model for dust, namely \textbf{d1s1}, but applied on data simulated with the \textbf{d6s1} model. In such cases, incorrect dust modeling leads to residuals in the clean CMB maps. 

We then construct the log-likelihood for $r$ assuming dust to be well modeled by \textbf{d1s1} using the noise covariance matrix in Eq.~(\ref{chi2}), $\textbf{N}_{\ell, \ell}$, obtained from simulations without frequency decorrelation in the dust emission. Such a covariance matrix does not incorporate the variance arising from the dust SED decorrelation so that the bias on $r$ appears with a high significance, which is precisely the effect we want to study. 

After a large number of realizations of \textbf{d6s1}, we see a distribution that shows the large spread in the possible values of $r$ which would be incorrectly considered a measure of high significance because we assumed a simple model for dust. We use the same scheme for all of our instrument configurations (from a classical imager to a bolometric interferometer with a number of sub-bands) so that we can explore if the extra spectral information provided by BI allows us to identify if the ``clean'' CMB maps after component separation are indeed clean or are contaminated by dust residuals.

    \subsection{Results}    
    \label{sec_results}   


\subsubsection{Reconstruction of the tensor-to-scalar ratio, $r$}
\label{sec:reconstruction_r_fgbuster}
\begin{figure*}[h!]
        \centering  
        \includegraphics[width=11cm]{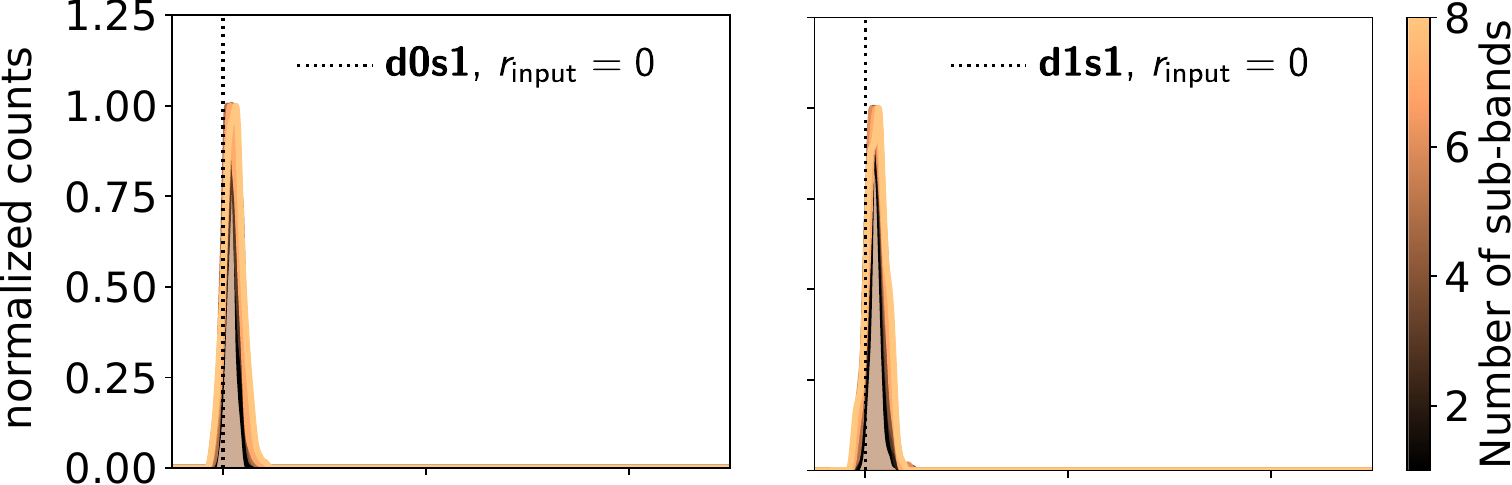} \\
        \includegraphics[width=11cm]{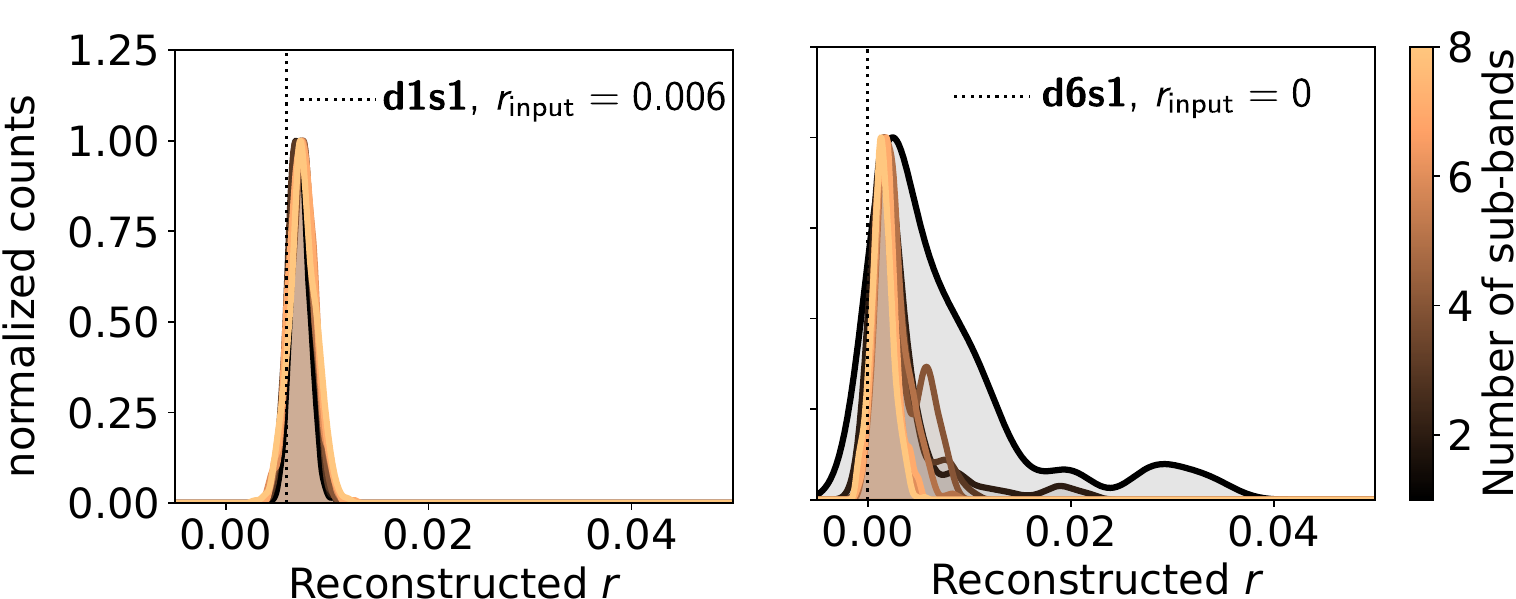}
        \caption{\label{fig:results_d0s0_r}Normalized histograms of the maximum likelihood values of $r$ as a function of the number of sub-bands. \textit{Top-left}: model \textbf{d0s0} with $r_\mathrm{input}=0$. \textit{Top-right}: model \textbf{d1s1} with $r_\mathrm{input}=0$. \textit{Bottom-left}: model \textbf{d1s1} with $r_\mathrm{input}=0.006$. \textit{Bottom-right}: model \textbf{d6s1} with $\ell_\mathrm{corr}=10$ and $r_\mathrm{input}=0$.}
    \end{figure*}

    Here we discuss the results of our FGBuster simulations in terms of the reconstruction of the tensor-to-scalar ratio, $r$. The performance in terms of foreground reconstruction is discussed in Appendix~\ref{app:reconstruction_foregrouds} (FGBuster simulations), whereas the Commander simulations are presented in Appendix~\ref{app:commander_results}.
    \\
    \\
    The four panels in Fig.~{\ref{fig:results_d0s0_r}} show the histograms of the maximum likelihood values of $r$ computed from Eq.~(\ref{chi2}) for each iteration of the Monte-Carlo chain. Each panel shows the result for one of the simulated sky models as a function of $n_\mathrm{sub}$. The \textit{top-left panel} shows the CMB with $r_\mathrm{input}=0$ and \textbf{d0s0} foregrounds.  The \textit{top-right panel} shows the CMB with $r_\mathrm{input}=0$ and \textbf{d1s1} foregrounds. The \textit{bottom-left panel} shows the CMB with $r_\mathrm{input}=0.006$ and \textbf{d1s1} foregrounds. The \textit{bottom-right panel} shows the CMB with $r_\mathrm{input}=0$ and \textbf{d6s1} foregrounds with $\ell_\mathrm{corr}=10$.

    The histograms are normalized to the maximum count value and smoothed with a kernel density estimator (KDE) of width equal to one-fourth of the standard deviation of the histogram. The histograms extend to negative $r$ because we compute the posterior likelihood over a range of $r$ that includes negative values in order to avoid a sharp truncation of the likelihood at $r=0$.
    
    A more detailed discussion of each of the four cases follows below.
    


    \paragraph{Top-left panel.} Here we have the CMB with $r_\mathrm{input}=0$ and \textbf{d0s0} foregrounds. In this case, the reconstructed $r$ does not depend on $n_\mathrm{sub}$ and there is a small bias due to an $E\rightarrow B$ modes leakage caused by the power spectra computation on a sky patch, where the spherical harmonics are no longer orthogonal. This bias could be mitigated by increasing the apodization radius of the mask at the expense of a smaller effective sky fraction ($<3\%$). This optimization, however, is outside the scope of the paper.

    \paragraph{Top-right panel.} Here we have the CMB with $r_\mathrm{input}=0$ and \textbf{d1s1} foregrounds. Also in this case we see that the reconstructed $r$ does not depend on $n_\mathrm{sub}$, even if the complexity of the dust emission is higher (the dust spectral index varies in the sky). However, here we observe a slightly larger bias in $r$ with respect to the \textbf{d0s0} case, caused by the aforementioned leakage and also by the difference in pixel size of the reconstructed spectral indices maps ($N_\mathrm{side}=8$) compared to the input sky ($N_\mathrm{side}=256$). 




    \paragraph{Bottom-left panel.} Here we have the CMB with $r_\mathrm{input}=0.006$ and \textbf{d1s1} foregrounds. This case is similar to the previous one, the only difference being the value of $r_\mathrm{input}$.
    
    \paragraph{Bottom-right panel.} Here we have the CMB with $r_\mathrm{input}=0$ and \textbf{d6s1} foregrounds fitted with the \textbf{d1s1} model. The histograms show that fitting with a model that does not account for frequency decorrelation produces distributions that are larger for smaller values of $n_\mathrm{sub}$. Also, the mean value of the reconstructed $r$ obtained from such distributions varies and becomes smaller as $n_\mathrm{sub}$ increases.
    \\



\begin{figure}[h!]
        \includegraphics[width=\columnwidth]{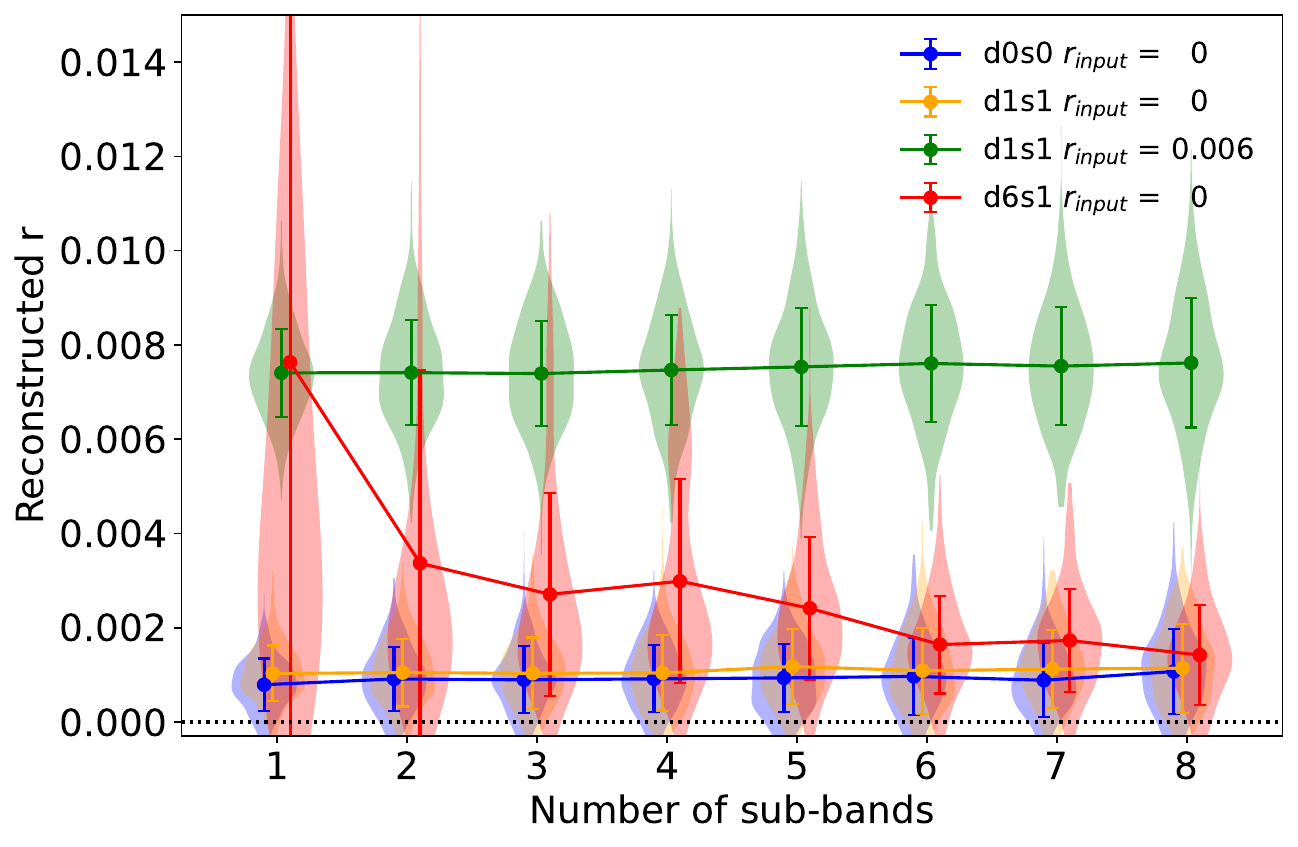} \\
        \caption{\label{fig:r_vs_nsub} Average maximum likelihood value of $r$ and standard deviation as a function of the number of sub-bands in the case of unaccounted dust frequency decorrelation (model \textbf{d6s1} with  $\ell_\mathrm{corr}=10$  and $r=0$) compared to two cases of no decorrelation (model \textbf{d1s1}): $r=0$ and $r=0.006$. On top of the average $r$ values and their standard deviation, we have overplotted the shape of the distribution as a ``violin plot''. Note that for the \textbf{d6} case the distribution is asymmetric for small $n_\mathrm{sub}$, so that the average is not centered on the distribution.}
    \end{figure}

    Fig.~\ref{fig:r_vs_nsub} shows the average $r$ and standard deviation computed from the histograms of Fig.~\ref{fig:results_d0s0_r} as a function of $n_\mathrm{sub}$. This result represents the range of $r$ from which we expect to sample our measurement when performing CMB observations.

    Note that since the error bar is the standard deviation, we assume it to be symmetrical. Moreover, in the \textbf{d6s1} case the histogram is unsymmetrical, and therefore the average $r$ is not centered with the distribution.
    
    The blue, orange, and green curves refer to the case in which we fit the same dust model used to simulate the input sky. In these three cases, the recovered $r$ does not depend on $n_\mathrm{sub}$, as one would expect for a detection not contaminated by foregrounds. The difference between the recovered $r$ with respect to $r_\mathrm{input}$ that we see in all three cases is caused by the $E\rightarrow B$ leakage and pixel size effects discussed above.

    The red curve refers to the case in which the input sky contains dust emission with frequency decorrelation while component separation was performed ignoring this feature, assuming the {\bf d1s1} model. In this case, the increase in the number of frequency maps provided by BI allows us to better constrain the spectral indices, thus reducing the bias as the number of sub-bands increases. On average, a classical imager (represented by $n_\mathrm{sub}=1$) would measure $r\sim 0.008$ while a bolometric interferometer would see this estimate reducing by increasing $n_\mathrm{sub}$. This indicates that the first value of $r$ is an artifact due to the presence of residual dust emission.
    
\begin{figure}[h!]
    \begin{center}
        \includegraphics[width=0.9\columnwidth]{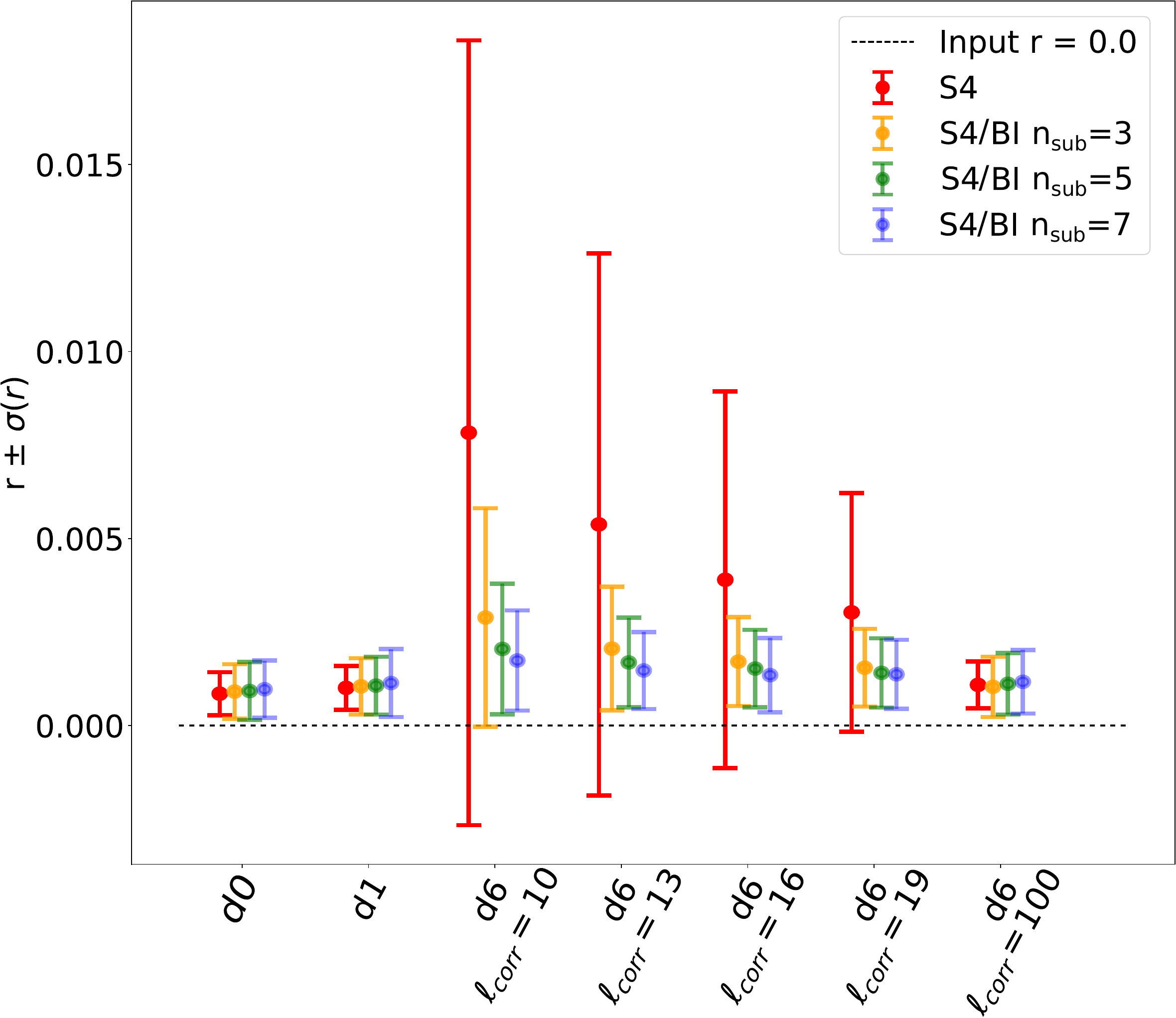} 
        \caption{\label{fig:results_d6s1_r}Summary of the average maximum likelihood value of $r$ and standard deviation for an input $r=0$ and all the simulated foreground models (\textbf{d0s0}, \textbf{d1s1} and several $\ell_\mathrm{corr}$ cases of \textbf{d6s1}). Note that we assume symmetric error bars.}
    \end{center}
    \end{figure}

    Finally, Fig.~{\ref{fig:results_d6s1_r}} shows a summary of the average $r$ and standard deviation for all the simulated dust models with $r_\mathrm{input}=0$, including various correlation lengths for the \textbf{d6s1} case: $\ell_{\mathrm{corr.}}=10,13,16,19,100$. For the sake of simplicity, we only show four instrument configurations: CMB-S4 and CMB-S4/BI with 3, 5, and 7 sub-bands. As one can see, the advantage of BI in diagnosing foreground residuals, and therefore decreasing the bias on $r$, is maintained even in the case of smaller levels of dust frequency decorrelation. As expected, in the limit of $\ell_{\mathrm{corr.}}=100$ the result is compatible with the case of a single modified black-body (model \textbf{d1s1}).

\subsubsection{Identifying foreground residuals on a single realization}
\label{classifier}
    We used machine learning to test the ability of BI to detect foreground residuals that may be present when the assumed foreground model is different from that describing the actual sky emission. That might occur, for example, if one assumes a \textbf{d1s1} model when the sky is described by a \textbf{d6s1} model. Therefore, we explore the possibility of classifying between ``contaminated'' and ``not contaminated'' cases that both end up producing the same average reconstructed $r$ for an imager (described by the case in which we do not split the physical band in sub-bands).

This ability is a key issue when an experiment detects a tensor-to-scalar ratio that is significantly different from zero. In this case, there is only one realization (i.e., the actual measurement) to understand whether there are unknown systematic effects biasing the value beyond the uncertainty set by the noise plus the known systematic effects.

We carried out this test by performing a machine learning classification based on a simple gradient-boosted decision tree (a \texttt{GradientBoostingClassifier} from the \texttt{scikit-learn} Python library\footnote{\protect\url{https://scikit-learn.org/}}) according to these steps:  

\begin{enumerate}
    \item Produce 500 sky realizations with $r=0.006$\footnote{The value of $r=0.006$ was chosen so that the average reconstructed $r$ matched the bias that would be obtained from a map with CMB with $r=0$ and \textbf{d6s1} foregrounds removed assuming a \textbf{d1s1} model with a single reconstructed sub-band(see Fig.~\ref{fig:r_vs_nsub})} in which the sky is generated with \textbf{d1s1} and fitted with the same model (we call this dataset \textbf{d1-d1}). This dataset is labeled as ``clean''; 
    \item Produce 500 simulations with $r=0$, in which the sky is generated with \textbf{d6s1} ($\ell_\mathrm{corr}=10$) and fitted with \textbf{d1s1} (we call this dataset \textbf{d6-d1}). This dataset is labelled as ``contaminated''; 
    \item \label{item:train_set} For each simulation, and for each value of $n_\mathrm{sub}$, calculate a normalized reconstructed $r$ and its uncertainty normalized by what is found with $n_\mathrm{sub}=1$, expressed by 
    the following two quantities: $\rho(n_\mathrm{sub}) = r(n_\mathrm{sub}) / r(n_\mathrm{sub}=1)$ and $\sigma_\rho(n_\mathrm{sub}) = \sigma(r(n_\mathrm{sub})) / r(n_\mathrm{sub}=1)$  (``training'' dataset), both with ``clean'' or ``contaminated'' label, depending on the model used as an input. These quantities are those that discriminate whether we have foreground residuals or not. If $\rho \neq 1$, it means that the detection depends on the number of sub-bands and, therefore, is likely to be affected by foreground residuals;
    \item Train the network with 250 \textbf{(d1s1, $r=0.006$)} and 250 \textbf{(d6s1, $r=0$)} randomly selected realizations from the training dataset (using 100 cross-validation subsets); 
    \item \label{item:predict_set} Calculate $\rho(n_\mathrm{sub})$ and $\sigma_\rho(n_\mathrm{sub})$ for the remaining 250 \textbf{(d1s1, $r=0.006$)} and 250 \textbf{(d6s1, $r=0$)} simulations (``test'' dataset);
    \item Feed the trained network with the values calculated in step \ref{item:predict_set} to test its ability to classify the simulations as ``clean'' (constant $\rho(n_\mathrm{sub})$) or ``contaminated'' (variable $\rho(n_\mathrm{sub})$\textbf{)}.
\end{enumerate}

\begin{figure}
    \centering
    \resizebox{\hsize}{!}
    {\includegraphics{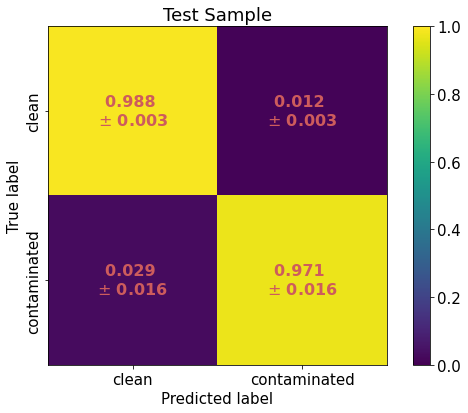}}
    \caption{Confusion matrix representing our ability to classify between our simulated data sets with dust frequency decorrelation (contaminated) or without (clean) using the measurements of $r$ as a function of $n_{sub}$. We observe that the fraction of false negatives (``contaminated'' data set incorrectly classified as ``clean'') is close to zero.}
    \label{fig:ML}
\end{figure}

The result of this procedure is the so-called ``confusion matrix'', i.e., a matrix that compares the results from the classification predicted by the algorithm with the true one as shown in Fig.~\ref{fig:ML}. The performance of our classifier is as follows (we adopted the convention ``clean=negative'' and ``contaminated=positive''):
\begin{itemize}
    \item True negative rate very close to 1, indicating that the realizations with no dust residuals (dataset {\bf d1-d1} with $r=0$ and $r=0.006$) displayed a constant ratio $\rho(n_\mathrm{sub})$ and were correctly classified as ``clean'';
    \item True positive rate very close to 1, indicating that the realizations with dust residuals (dataset {\bf d6-d1} with $r=0$), displayed a variable ratio $\rho(n_\mathrm{sub})$ and were correctly classified as ``contaminated'';
    \item Low false negative rate of $2.9\%\pm 1.6\%$, indicating a very low percentage of realizations with dust residuals that were wrongly classified as ``clean''. This is a very important figure of merit that we want to minimize;
    \item Low false positive of $1.2\% \pm 0.3\%$, indicating a very low percentage of realizations without dust residuals that were wrongly classified as ``contaminated''.
\end{itemize}

Such high classification performance demonstrates that BI, with its capability to measure $r$ in several sub-bands, is a promising solution to identify residuals in the clean CMB maps arising from LOS frequency decorrelation in the dust emission. In such a case, a classical imager lacks the frequency resolution to identify this contamination, leading to a systematic uncertainty in the reconstructed $r$ that is well above the target sensitivity of \mbox{CMB-S4}.

\section{Conclusions}
\label{sec_conclusions}
In this paper, we have shown how bolometric interferometry (BI) has the potential to detect systematic effects caused by interstellar dust in CMB polarization measurements when LOS frequency decorrelation is present in dust emission and it is not accounted for in parametric component separation algorithms. 

We know that there are ways for imagers to mitigate the problem of not precisely knowing the foreground emission, for example, through cross-checking with different component separation methods, such as blind ones \citep{Aumont_2007}, or codes based on the moment expansion \citep{Chluba_2017, Vacher_2022} which might be less sensitive to incorrect foreground modeling. However, in this paper, we propose a new approach based on a different instrument architecture called bolometric interferometry (BI).  An instrument based on BI can be used as an independent verification to future claims of a \mbox{$B$-mode} detection by exploiting the superior purity of the $r$ measurement made possible by the increased spectral resolution.

We have carried out simulations with two component separation codes (FGBuster, discussed in the main text, and Commander, discussed in Appendix~\ref{app:commander}), reconstructing the tensor-to-scalar ratio, $r$, from simulated skies containing CMB, synchrotron and dust emission, and instrumental noise. For the dust emission, we used three models of increasing complexity, one of which contains frequency decorrelation.

We compared two instrument models, CMB-S4 and \mbox{CMB-S4/BI}, the latter being a modified version of CMB-S4 that accounts for the possibility of splitting each physical frequency band in a variable number of sub-bands that can be chosen during data analysis. This feature, which is unique to BI, allows us to assess whether a measurement of $r$ is biased by dust emission residuals or not. While a Fourier-transform spectrometer can be used to increase spectral resolution, it would suffer from a noise penalty compared to BI because it cannot observe all frequencies simultaneously.

Our results are consistent for the two codes and show that with no frequency decorrelation, both instruments yield the same performance (the final precision and systematic uncertainty on $r$ are similar). If decorrelation is present, and it is not accounted for in component separation, then an imager like CMB-S4 would measure a biased value of $r$. This bias can be reduced with \mbox{CMB-S4/BI} by reanalyzing the same data after splitting the band into an increasing number of sub-bands. 

The decrease of the measured $r$ with the number of sub-bands, $n_\mathrm{sub}$, clearly indicates the presence of a dust-induced systematic effect, given that without dust residuals the detected $r$ does not change with $n_\mathrm{sub}$. In such a situation, a classical imager would have no means of classifying the measurement as ``clean'' or ``biased''.

We tested the ability to detect biased $r$ measurements also using a machine learning approach, and we verified that assessing the variation of the $r$ measurement versus $n_\mathrm{sub}$ allowed us to classify clean and biased measurements with a rate $\gtrsim 97\%$.


Future developments will test this technique in more realistic situations (representative noise, inclusion of optical effects, and uncertainty on the knowledge of the instrumental spectral response), assess the performance with various dust models, and explore new techniques of component separation, allowing us to separate signals taking into account instrumental effects in a more comprehensive and representative way.

\begin{acknowledgements}
QUBIC is funded by the following agencies. France: ANR (Agence Nationale de la
Recherche) contract ANR-22-CE31-0016, DIM-ACAV (Domaine d’Intérêt Majeur-Astronomie et Conditions d’Apparition de la Vie), CNRS/IN2P3 (Centre national de la recherche scientifique/Institut national de physique nucléaire et de physique des particules), CNRS/INSU (Centre national de la recherche scientifique/Institut national et al de sciences de l’univers). Italy: CNR/PNRA (Consiglio Nazionale delle Ricerche/Programma Nazionale Ricerche in Antartide) until 2016, INFN (Istituto Nazionale di Fisica Nucleare) since 2017.  Argentina: MINCyT (Ministerio de Ciencia, Tecnología e Innovación), CNEA (Comisión Nacional de Energía Atómica), CONICET (Consejo Nacional de Investigaciones Científicas y Técnicas).
 
 S. Paradiso acknowledges support from the Government of Canada's New Frontiers in Research Fund, through grant NFRFE-2021-00595.

 The authors want to thank Alexandre Boucaud for valuable advices about Machine Learning.
\end{acknowledgements}

\bibliographystyle{aa} 
\bibliography{biblio,qubic} 

\appendix

\section{Reconstruction of foregrounds parameters}
\label{app:reconstruction_foregrouds}



In our paper, we focused on the reconstructed tensor-to-scalar ratio, $r$, as it is the main quantity of interest. The level of systematic uncertainties in the reconstructed $r$, however, depends on the reconstructed foreground spectral parameters and distributions. Thus, in this appendix we focus on the distribution of the foregrounds spectral indices after component separation. 

\begin{figure}[h!]
    \resizebox{\hsize}{!}
	{\includegraphics[width=9cm]{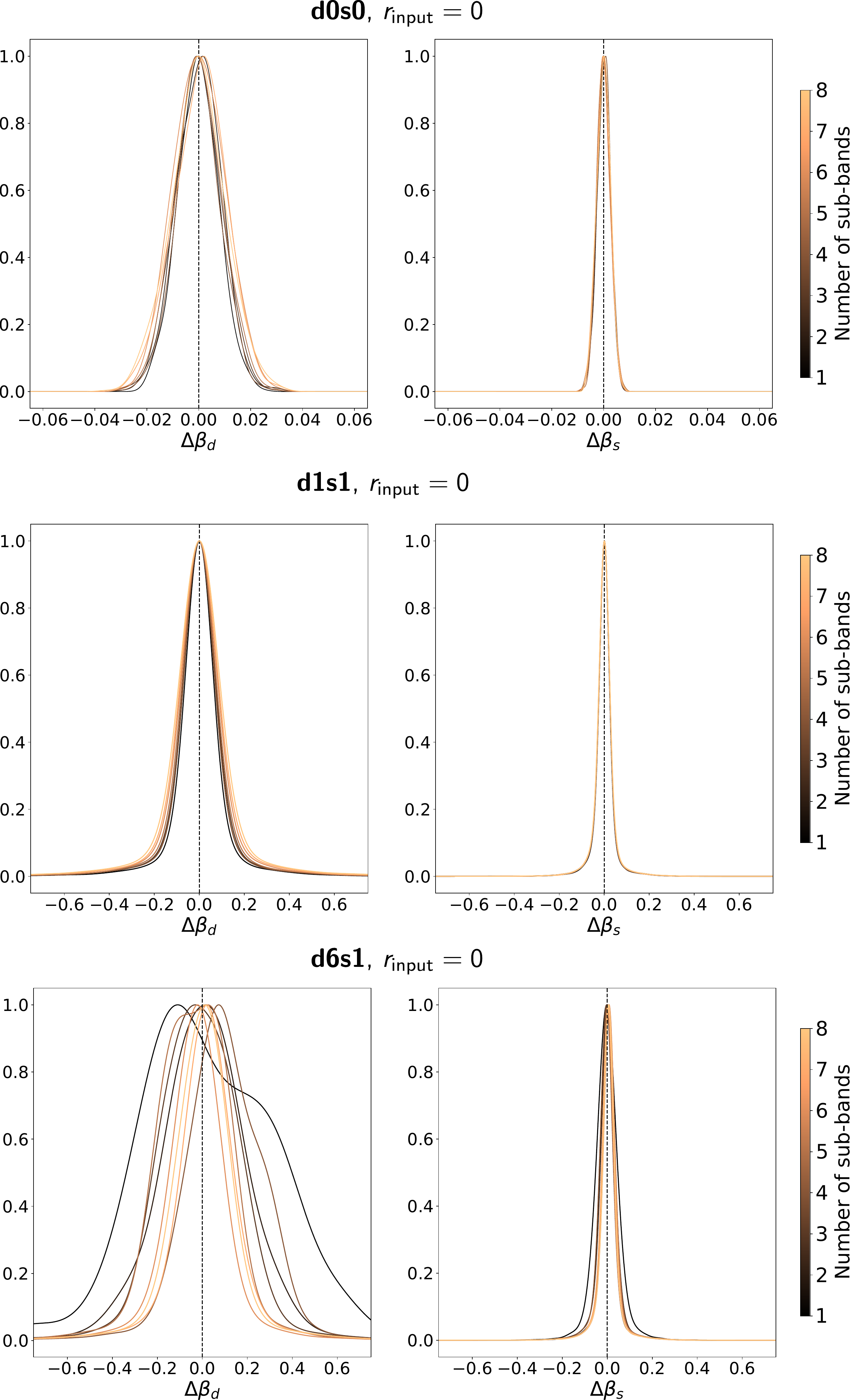}}
    \caption{Reconstruction of foregrounds spectral indices. \textit{Top}: model \textbf{d0s0}, $r_\mathrm{input}=0$. \textit{Middle}: model \textbf{d1s1}, $r_\mathrm{input}=0$. \textit{Bottom}: model \textbf{d6s1}, $r_\mathrm{input}=0$.}
    \label{fig:histograms_foregrounds}
\end{figure}

In Fig.~\ref{fig:histograms_foregrounds} we show the normalized histograms of the difference between the reconstructed and input dust and synchrotron spectral indices, $\Delta\beta_\mathrm{d}$, $\Delta\beta_\mathrm{s}$ for the following three models: \textbf{d0s0} (top row), \textbf{d1s1} (middle row), \textbf{d6s1} (bottom row), all with \mbox{$r_\mathrm{input}=0$}. Each histogram does not correspond to a particular pixel but contains values from the sky patch. 

In the case of \textbf{d0s0}, the model assumes a constant spectral index all over the sky. Therefore we expect unbiased estimates with a standard deviation related to the noise in the input frequency maps. The results shown in the top row of Fig.~\ref{fig:histograms_foregrounds} confirm this expectation as we observe no bias on the reconstructed spectral indices. We note that the standard deviation slightly increases with the number of sub-bands, $n_\mathrm{sub}$, because of the slight sub-optimality inherent to spectral-imaging \cite[parametrized by $\varepsilon$ in Eq.~\ref{sigma}, see][]{2020.QUBIC.PAPER2}. 



When spectral indices vary across the sky, as in \textbf{d1s1}, we expect biases in the reconstructed spectral indices because we only reconstruct the spectral indices on relatively large sky pixels ($N_\mathrm{side}=8$), while the input sky was simulated with spectral indices that vary among smaller pixels ($N_\mathrm{side}=256$). Consequently, averaging multiple spectral indices in large pixels introduces a bias to the reconstructed spectral index. This bias is responsible for foreground residuals in the CMB maps obtained after component separation and produces the bias on $r$ observed in Figs.~\ref{fig:results_d0s0_r} and ~\ref{fig:r_vs_nsub}.

This is shown in the middle row of Fig.~\ref{fig:histograms_foregrounds}. The bias due to spatial decorrelation appears as an enlarged spread of the distribution with respect to the \textbf{d0s0} case (notice the increased scale of the $x$-axis in the middle row compared to the top row). Also in this case we observe an increase in standard deviation with $n_\mathrm{sub}$ caused by the sub-optimality related to spectral imaging.

Finally, in the case of frequency decorrelation in the dust emission ({\bf d6s1} model), spectral indices are no longer an accurate description of the dust spectral behavior. As a result, if we reconstruct $\beta_\mathrm{d}$ using a {\bf d1s1} model we expect a much larger bias. Note that the comparison between input and reconstructed spectral indices is done using the template map of $\beta_{\mathrm{d}}$ \citep{Planck_15} that was used as an input for the {\bf d6s1} model. It is clear however that in the case of {\bf d6s1}, the comparison between the input and recovered spectral indices is less meaningful than in the simpler models. In this case, one is more interested in the residuals found in the "clean" CMB maps, discussed in the main text of this article. In this case, the increase in spectral resolution provided by spectral imaging supplies extra information, allowing us to reduce this bias. 
This is confirmed by the results shown in the bottom row of Fig.~\ref{fig:histograms_foregrounds}. First, we see a much larger spread in the histograms compared to the other two cases.  Second, we see that the spread reduces significantly by increasing $n_\mathrm{sub}$. In this case the benefit from spectral imaging more than balances the sub-optimality effect and allows us to reduce the bias on the reconstructed spectral index which then reduces the bias on $r$, as shown in Fig.~\ref{fig:r_vs_nsub}.

\section{Simulations with Commander}
\label{app:commander}

    \subsection{Simulation pipeline}
    \label{app:commander_pipeline}
    
We describe here the simulation pipeline for the analysis performed using the Commander code \citep{eriksen_2006,eriksen_2008}.
We generated 100 CMB power spectra using CAMB \citep{Lewis:1999bs} from the set of cosmological parameters shown in Table \ref{tab:1}.

\begin{table}[h!]
    \caption{\label{tab:1}Set of cosmological parameters from the \textit{CAMB Python example notebook} in the CAMB documentation (\protect\url{https://camb.readthedocs.io/en/latest/CAMBdemo.html}).
    }
    \renewcommand{\arraystretch}{2}
    \begin{center}
        \begin{tabular}{p{2cm} m{2cm}}
\hline
\hline
\makecell[c]{$\mathrm{H}_0$} & \makecell[c]{67.5}\\
\makecell[c]{$\Omega_b h^2$} & \makecell[c]{0.022}\\
\makecell[c]{$\Omega_c h^2$} & \makecell[c]{0.122}\\
\makecell[c]{$\Omega_K$} & \makecell[c]{0}\\
\makecell[c]{$m_\nu$} & \makecell[c]{0.06}\\
\makecell[c]{$\tau$} & \makecell[c]{0.06}\\
\makecell[c]{$A_s \times 10^{-9}$} & \makecell[c]{2}\\
\makecell[c]{$n_s$} & \makecell[c]{0.965}\\
\hline

\end{tabular}
\end{center}
\end{table}

\begin{table}[h!]
    \caption{\label{tab:2}Parameters used for analyzing simulations with Commander.}
    \renewcommand{\arraystretch}{2}
    \setlength{\tabcolsep}{1pt}
    \begin{center}
        \begin{tabular}{p{5cm} m{2cm}}
            \hline
            \hline
            Number of CMB realizations\dotfill & \makecell[c]{100}\\
            Map $N_\mathrm{side}$\tablefootmark{a} \dotfill	& \makecell[c]{64} \\
            Multipole range\tablefootmark{b} \dotfill	&\makecell[c]{21--128} \\
            $\Delta\ell$\dotfill &\makecell[c]{\phantom{0}{35}} \\
            Input $r$\dotfill & \makecell[c]{\phantom{00}{0}} \\
            Residual lensing fraction\tablefootmark{c} \dotfill & \makecell[c]{\phantom{000}100\%} \\
            Sky fraction [\%]\dotfill & \makecell[c]{\phantom{0000}3\%} \\
            \makecell[l]{Sky patch center\\ \mbox{}[Equatorial coord.]\dotfill}\dotfill & 
                \makecell[c]{$\text{RA}=0^\circ$\\\phantom{1}$\text{DEC}=-57^\circ$\rule[-2.2ex]{0pt}{0pt}}\\
            FWHM\dotfill & \makecell[c]{1$^\circ$}\\
            \hline
            \multicolumn{2}{l}{\makecell[l]{
              \tablefoot{\\
               \tablefoottext{a}{Limited by computational time.}
               \\
               \tablefoottext{b}{Limited by $N_\mathrm{side}=64$.}
               \\
               \tablefoottext{c}{The value of 100\% means that all the lensing signal was left.}}
              }}
        \end{tabular}
    \end{center}
\end{table}

We smoothed both the CMB and foreground signals with a Gaussian beam with FWHM of 1$^\circ$ and applied the HEALPix pixel window function at $N_\mathrm{side}=64$.
The only model used to generate the foreground is the $\mathbf{d6s1}$ described in Sect. \ref{sec_simulated_sky_model}, in particular setting the dust correlation length to $\ell_\mathrm{corr}=10$.

For this test we considered a circular patch covering 3\% of the sky, centered on the QUBIC observation field ---corresponding to RA = $0^\circ$ and DEC = $-57^\circ$. 
We made this choice in order to be consistent with an already-existing BI experimental set-up, after observing that such a sky region is reasonably close to the CMB-S4 one
---considered throughout the analysis in the main text. We note that the foreground contamination is going to differ in the two pipelines, as well as the CMB realization,
leading to a slightly different estimate of $r$. Nevertheless, we still expect the final posterior distribution to be compatible within the instrumental uncertainty, the component separation residual 
contamination, and the additional statistical uncertainty arising from the different CMB realizations. However, the latter contribution is reduced by $\sqrt{N}$, where
$\mathrm{N}=100$ is the number of CMB realizations considered in the analysis.

For computational reasons only four configurations have been studied.  They are \mbox{CMB-S4/BI} with 1, 3, 5, and 7 sub-bands.

For each simulated sky map, we generated a second version by taking the same CMB, synchrotron, and dust realization, and a different Gaussian noise realization.
The analysis chain is the same as outlined in Sect. \ref{sec_simulation_pipeline} except that we use an apodization radius of 4.6$^\circ$ instead of 4$^\circ$.
We performed the component separation sampling the amplitudes $\mathrm{a}_{\mathrm{CMB}}, \mathrm{a}_{\mathrm{s}}, \mathrm{a}_{\mathrm{d}}$ and the spectral indices $\beta_{\mathrm{s}}, \beta_{\mathrm{d}}$ by means of the following Gibbs chain:
\begin{align}
        \{\mathrm{a}_{\mathrm{CMB}},\mathrm{a}_\mathrm{s}, \mathrm{a}_\mathrm{d}\}^{i+1} &\leftarrow P(\mathrm{a}_{\mathrm{CMB}},\mathrm{a}_\mathrm{s}, \mathrm{a}_\mathrm{d} \,|\, \beta_\mathrm{s}^i,\beta_\mathrm{d}^i,\mathrm{d})\tag{7a}\label{eq:gibbs_chain_1}\\
        \notag\\
        \beta_\mathrm{s}^{i+1} &\leftarrow P(\beta_\mathrm{s} \,|\, \mathrm{a}_{\mathrm{CMB}}^{i+1},\mathrm{a}_\mathrm{s}^{i+1}, \mathrm{a}_\mathrm{d}^{i+1},\beta_\mathrm{d}^i,\mathrm{d})\tag{7b}\label{eq:gibbs_chain_2}\\
        \notag\\
        \beta_\mathrm{d}^{i+1} &\leftarrow P(\beta_\mathrm{d} \,|\, \mathrm{a}_{\mathrm{CMB}}^{i+1},\mathrm{a}_\mathrm{s}^{i+1}, \mathrm{a}_\mathrm{d}^{i+1},\beta_\mathrm{s}^{i+1},\mathrm{d})\tag{7c}\label{eq:gibbs_chain_3}\,.
\end{align}

The spectral indices are sampled at $N_\mathrm{side}=8$ as for the \mbox{FGBuster} pipeline.
We generated 1000 MCMC samples for each input sky realization and discarded the first 100 samples as burn-in.
The two noise uncorrelated versions of the same sky realization are associated with two parallel sampling chains.
We compute the cross-spectra between these two parallel chains, iteration by iteration, to collect a set of 900 spectra for each CMB realization considered in the analysis. We also average all of the sampled maps produced in a single chain into a mean map, and for every couple of parallel chains we compute the cross-spectrum between the two mean maps.

After the component separation, we compute the likelihood function for the cross spectrum of each mean map, exploiting the sample-based noise covariance matrix obtained by all the power spectra from the corresponding sampling chain.

    \subsection{Results}
    \label{app:commander_results}

\begin{figure}[ht]
	\includegraphics[width=9cm]{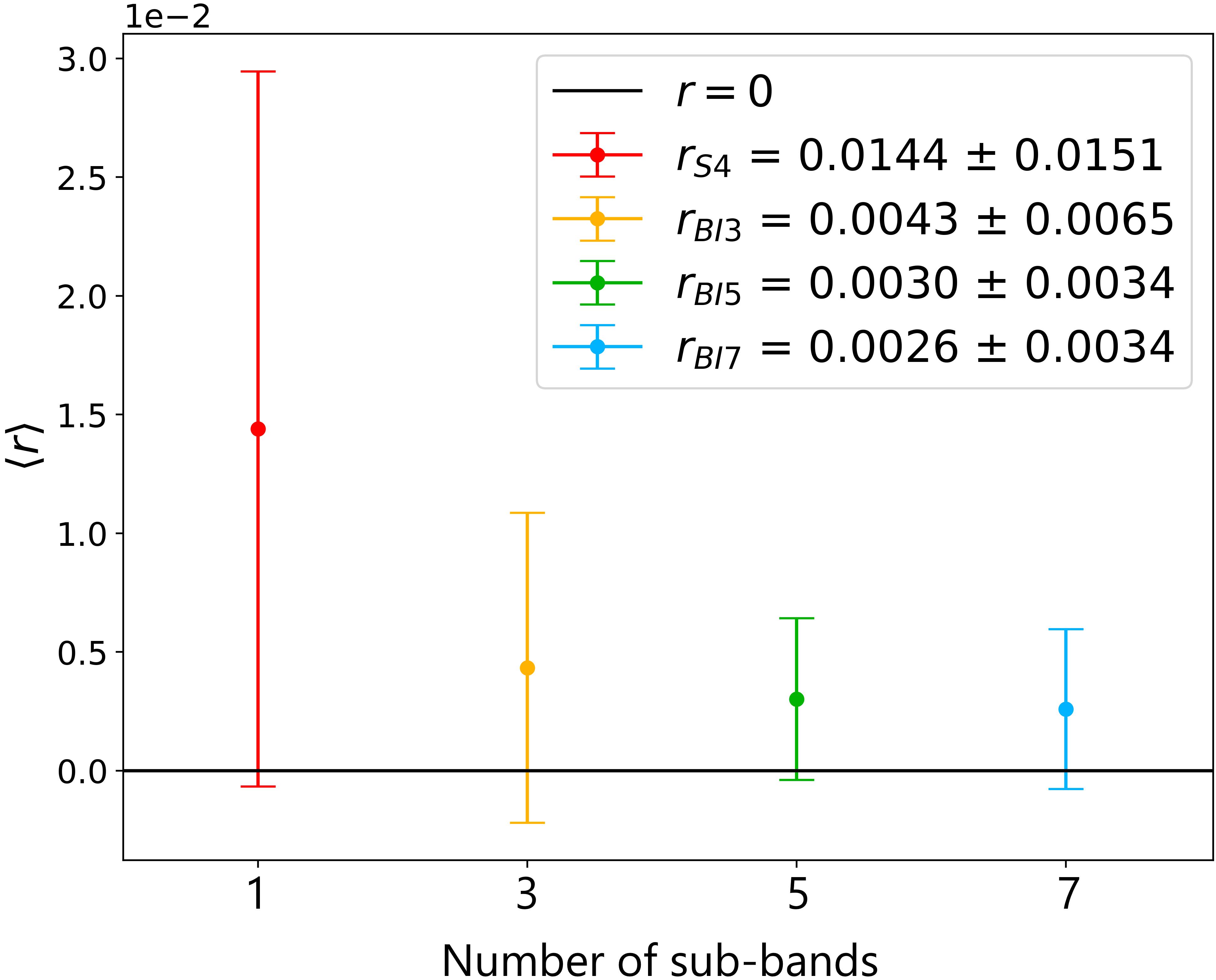}
        \caption{\label{fig:commander_r0}Mean and standard deviation of the best fit distributions
        obtained with Commander, using the \textbf{d6s1} model with $\ell_\mathrm{corr}=10$ and $r=0$.}
\end{figure}

    From the probability density functions of the model parameters obtained with Commander, we find that the upper limit to the estimation of a single realization of $r$ is reduced with the number of sub-bands, as shown in Figure \ref{fig:commander_r0}. The $r$ bias and $\sigma(r)$ are greater than the FGBuster results due to the marginalization over the foreground components.

\begin{figure}[ht]
    \hspace{-0.1cm}
	\includegraphics[scale=0.2]{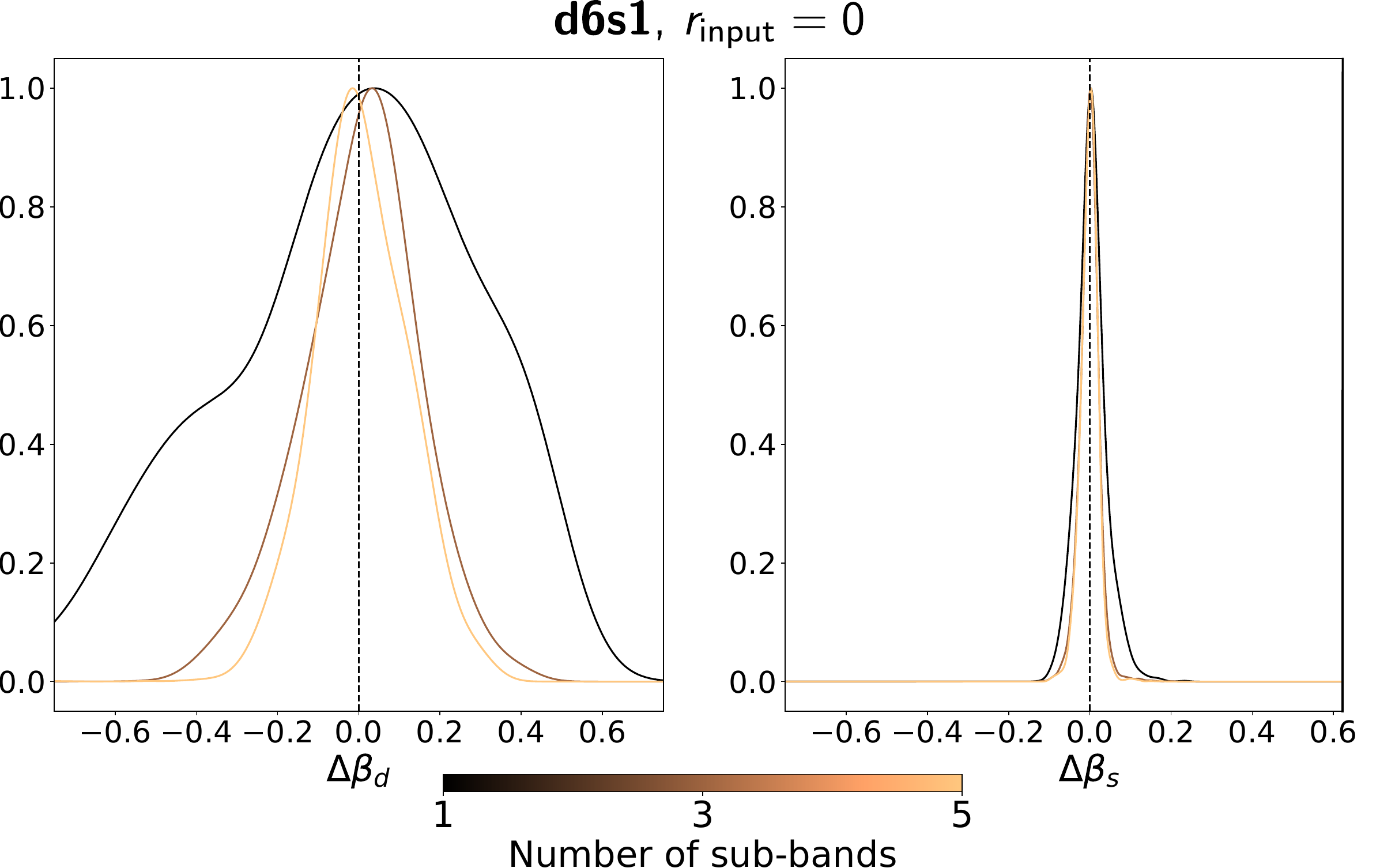}
        \caption{\label{fig:commander_indices_all_pixels}: Reconstruction of foreground spectral indices for the \textbf{d6s1} model with the Commander pipeline.}
\end{figure}

Increasing the number of sub-bands also reduces the standard deviation of the marginalized posterior distributions of the standard deviation of the spectral indices for all pixels. Figure \ref{fig:commander_indices_all_pixels} shows the comparison between the reconstructed dust and synchrotron spectral indices. As in Appendix~\ref{app:reconstruction_foregrouds} we compare reconstructed spectral indices with the input ones, from~\cite{Planck_15} for 1, 3, and 5 sub-bands on all pixels. Because of frequency decorrelation, the spectral indices residuals for dust are not as meaningful as the distribution of reconstructed $r$ shown in Fig.~\ref{fig:commander_r0}. This analysis has not been performed for the 7 sub-band configuration results because of data storage issues. Here a single $\Delta \beta$ from the plotted distributions represents the difference between the mean value of the marginal distribution on a single pixel for a given sky realization and the template value in the same pixel from the model. These results are in agreement with the FGBuster simulations.


\end{document}